\documentclass[fleqn,usenatbib]{mnras}
\usepackage{newtxtext,newtxmath}
\usepackage[T1]{fontenc}
\usepackage{ae,aecompl}
\usepackage{graphicx}	% Including figure files
\usepackage{amsmath}	% Advanced maths commands
\usepackage{array}
\usepackage{multirow}
\newcolumntype{H}{>{\setbox0=\hbox\bgroup}c<{\egroup}@{}}
\title[Stellar populations in AGNs]{Stellar populations in local AGNs: evidence for enhanced star formation in the inner 100\,pc}

\author[Dahmer-Hahn et al.]{L. G. Dahmer-Hahn $^{1,2}$\thanks{E-mail: lhahn@lna.br}, R. Riffel$^3$, A. Rodr\'iguez-Ardila$^{1,4}$, R. A. Riffel$^5$, \newauthor T. Storchi-Bergmann$^3$, M. Marinello$^1$, R. I. Davies$^6$, L. Burtscher$^7$, D. Ruschel-Dutra$^8$, \newauthor D. J. Rosario$^9$
\\
$^1$Laborat\'orio Nacional de Astrof\'isica/MCTIC, Rua dos Estados Unidos, 154, Bairro das Na\c{c}\~oes, Itajub\'a, MG, Brazil\\
$^2$Shanghai Astronomical Observatory, Chinese Academy of Sciences, 80 Nandan Road, Shanghai 200030, China\\
$^3$Instituto de F\'isica, Universidade Federal do Rio Grande do Sul, Av. Bento Gon\c{c}alves 9500, 91501-970 Porto Alegre, RS, Brazil\\
$^4$Instituto Nacional de Pesquisas Espaciais. Divisão de Astrof\'isica. Avenida dos Astronautas 1758. São Jos\'e dos Campos, 12227-010, SP, Brazil\\
$^5$Departamento de F\'isica, Centro de Ci\^encias Naturais e Exatas, Universidade Federal de Santa Maria, 97105-900, Santa Maria, RS, Brazil\\
$^6$Max-Planck-Institut f\"ur Extraterrestrische Physik, Postfach 1312, 85741, Garching, Germany\\
$^7$Leiden Observatory, PO Box 9513, 2300 RA, Leiden, The Netherlands\\
$^8$Departamento de F\'isica, Universidade Federal de Santa Catarina, P.O. Box 476, 88040-900, Florian\'opolis, SC, Brazil\\
$^9$Centre for Extragalactic Astronomy, Department of Physics, Durham University, South Road, DH1 3LE, Durham, UK
}

\date{Accepted XXX. Received YYY; in original form ZZZ}

\pubyear{2015}

\begin{document}
\label{firstpage}
\pagerange{\pageref{firstpage}--\pageref{lastpage}}
\maketitle

\begin{abstract}

In modern models and simulations of galactic evolution, the star formation in massive galaxies is regulated by an {\it ad hoc} active galactic nuclei (AGN) feedback process. However, the physics and the extension of such effects on the star formation history of galaxies is matter of vivid debate.  In order to shed some light in the AGN effects over the star formation, we analyzed the inner 500$\times$500\,pc of a sample of 14 Seyfert galaxies using GMOS and MUSE integral field spectroscopy. We fitted the continuum spectra in order to derive stellar age, metallicity, velocity and velocity dispersion maps in each source. After stacking our sample and averaging their properties, we found that the contribution of young SP, as well as that of AGN featureless continuum both peak at the nucleus. The fraction of intermediate-age SPs is smaller in the nucleus if compared to outer regions, and the contribution of old SPs vary very little within our field of view (FoV). We also found no variation of velocity dispersion or metallicity within our FoV. Lastly, we detected an increase in the dust reddening towards the center of the galaxies. These results lead us to conclude that AGN phenomenon is usually related to a recent star formation episode in the circumnuclear region.

\end{abstract}


\section{Introduction}

Active Galactic Nuclei (AGN) are among the most energetic phenomena in the universe, with integrated bolometric luminosities up to 10$^{48}$\,ergs\,s$^{-1}$ \citep[e.g.][]{Koratkar&Blaes99,Bischetti+17}. The current paradigm of an AGN is that of a supermassive black hole (SMBH) being fed by an accretion disc. Around this structure, there is a dust and gas torus, and depending on the orientation between the clumpy torus and the observer, the SMBH may be visible or obscured  \citep{Ferland&Netzer83,Zier&Biermann02,RamosAlmeida&Ricci17,Audibert+17,Izumi+18}. 
\par
It is well established that the mass of the SMBH scales with the velocity dispersion ($\sigma$) of the host galaxy bulge \citep[in galaxies with classical bulges][]{Ferrarese&Merritt00,Gebhardt+00,Kormendy&Ho13}. Another important relation that helps shedding some light on this matter is the fact that both star formation rate (SFR) and AGN luminosity reach their peak around z$\sim$2 \citep{Silverman+08,Madau&Dickinson14}. However, if these correlations are incidental or if there are physical mechanisms behind them is still under debate.
\par
In this context, there is a strong discussion whether AGN activity is correlated with the star formation of the host galaxy, since it could help explaining the above results. Some authors \citep{StorchiBergmann+00,Kauffmann+03,Lutz+10} argue that in order to trigger the AGN, gas would neet to lose angular momentum, which would in turn cause a density enhancement, thus triggering a starburst phase. On the other hand, other authors \citep{Leslie+16,Ellison+16} suggest that the gas outflows caused by the AGN would remove gas from the inner regions of the galaxy, quenching star formation. Moreover, there are some authors \cite[e.g.][]{Stanley+17} that argue that the AGN has no effect on the star formation within the host galaxy, either because they are not powerful enough, or because both mechanisms mentioned earlier would cancel each other out.
\par
Important factors regarding this relation are the spatial and temporal scale analyzed. \citet{Hopkins&Quataert10} performed multi-scale and multi-resolution hydrodynamic simulations to study the inflow of gas leading to AGN activity. They found that in this pre-activity scenario, there is a rise in the star formation, due to the gas funneled to the center both triggering the AGN as well as forming stars. Also, they found that the nuclear star formation is more tightly coupled to AGN activity than the global star formation. On the other hand, \citet{Zubovas&Bourne17} analyzed through simulations how the activated AGN affects turbulent gas spheres.  They found that, after a certain critical AGN luminosity (which they determined to be 5$\times$10$^{46}$\,ergs\,s$^{-1}$), AGN activity mainly acts in removing gas from the inner $\sim$1\,kpc, thus quenching star formation in this region. 
\par
For the above reason, it is then fundamental to probe the impact of AGNs over small and large scales, as well as determine the time scale in which this co-evolution occurs. Constraining the time scale of this impact can be achieved by mapping the age of the stellar populations (SPs). If young SPs are present in large quantities, AGN fueling is contemporary with star formation; if instead intermediate age SPs dominate, fueling is driven by a post-starburst phase; lastly, finding only old stars would imply that gas inflow to the AGN is efficient and star formation is not related to nuclear activity.
\par
In the past decades, many studies were published analyzing the stellar populations of galaxies hosting AGNs \citep[e.g.][]{StorchiBergmann+01, CF+04, Riffel+09,Morelli+13, Cai+20}. However, these analyses were usually based on integrated spectra, resulting in very limited spatial information. Nowadays, with integral field spectroscopy, it is possible to resolve these regions in large samples of galaxies, allowing studies focused on mapping stellar properties across the body of the galaxies. \citet{Sanchez+18} analyzed 98\,AGNs with MaNGA (Mapping Nearby Galaxies at Apache Point Observatory) optical datacubes, and found that their AGNs are mainly located in the so-called green valley, a transition region between the blue, star-forming galaxies and the red, quenched ones \citep[see][for a broader discussion]{Wyder+07,Martin+07,Salim+07}. This green valley region appears to be composed of galaxies experiencing a transition between star-forming and quenched galaxies. Also, \citet{Sanchez+18} reported that the quenching of the star formation happens inside-out, but they found no difference between AGN hosts and non-active galaxies in the green valley, suggesting that the AGN itself has no immediate impact over the stellar population of its host galaxy. In another project, \citet{Mallmann+18} analyzed 62 AGNs, also from the MaNGA survey, and reported that the fraction of young stellar populations in high-luminosity AGNs is higher in the inner regions (r$\leq$0.5\,R$_e$, where R$_e$ is the effective radius) when compared with the control sample of non-AGNs, but they found no difference between the stellar populations of the low-luminosity AGNs and the control sample. In another study focused in the stellar populations of AGN hosts, \citet{Lacerda+20} analyzed 867 galaxies from the CALIFA survey (Calar Alto Legacy Integral Field Area), with 34 of them classified as AGNs. They were able to confirm that AGNs hosts are found in the green valley in almost all analysed properties with bimodal distributions, with AGNs mainly acting in removing or heating the molecular gas, which would quench the star formation of the galaxy.
\par
Previous studies of individual AGN hosts in the NIR, on the other hand, found rings of stellar populations with ages between 0.1 and 2.0\,Gyr \citep[e.g.][]{RiffelRA+10,Riffel+11,StorchiBergmann+12,Schonell+17,luisgdh+19a,Diniz+19}. The authors also found that these regions are usually associated with rings of lower stellar $\sigma$, suggesting that the intermediate-age stellar populations were formed from low-velocity dispersion gas.
\par
It can be seen that the relationship between star formation and AGN activity is far from being fully understood. One point to mention is that most works so far lack enough spatial resolution to allow sampling the stellar populations at scales down to a a few hundred of parsecs, with also big enough samples to allow conclusions on the general properties of AGNs, rather than on individual sources.
\par
In this paper, we analyze 14 Seyfert galaxies with optical datacubes from GMOS and MUSE. We aim at mapping the stellar properties of these objects, in order to determine the impact of the AGN over the host galaxy evolution. Whereas previous results obtained with MaNGA and CALIFA have focused on a bigger field of view (FoV) but with lower spatial resolution, we focused on the inner 5\farcs0$\times$5\farcs0 of these galaxies, but with a much better spatial resolution (~1$^{\prime \prime}$ corresponding to 78 to 259\,pc).
\par
This paper is structured as follows: in Section~\ref{sec:sample}, we detail our sample, data reduction and treatment, and in Section~\ref{sec:method} we present the methods used to analyse the stellar properties. In Section~\ref{sec:results}, we show our results and discuss the sources individually, whereas in Section~\ref{sec:discussion} we discuss the properties of the sample as a whole. Lastly, the conclusions are presented in Section~\ref{sec:conclusions}.

\section{Sample}
\label{sec:sample}

Our sample was selected as an optical follow up of the NIR sample described by \citet{RiffelRA+18}, based on the Swift-BAT 60-month catalogue \citep{Ajello+12}, limited to galaxies with 14$-$195\,keV luminosities L$_X \geq 10^{41.5}$\,erg\,s$^{-1}$ and redshift z$\leq$0.015. The ultrahard (14-195\,keV) band of the Swift-BAT survey measures direct emission from the AGN rather than scattered or re-processed emission, and is much less sensitive to obscuration in the line of sight than soft X-ray or optical wavelengths. This allows a selection based only on the AGN properties. As additional criteria, the authors limited their sample to galaxies accessible from Gemini North ($-$30$^{\circ}\leq\delta\leq$73$^{\circ}$), with nuclei that are bright and point-like enough to serve as tip-tilt guide references, or those with natural guide stars in the AO patrol field. The galaxies were also selected for having been previously observed in the optical and with extended [O{\sc iii}]$\lambda$5007 emission confirmed in the literature.
\par
We cross-correlated the NIFS sample with Gemini Multi-Object Spectrographs (GMOS) archival data, as well as Multi Unit Spectroscopic Explorer (MUSE) processed archival data. After removing duplicates and galaxies with poor data quality, our final sample is composed of 10 GMOS and six MUSE datacubes.  Table~\ref{tab:sample} presents the resulting sample, as well as their main properties.
\par
For GMOS data, we followed the standard reduction process, which included trimming, bias subtraction, bad pixel and cosmic ray removal, extraction of the spectra, Gemini Facility Calibration Unit (GCAL) flat correction, twilight flat correction, wavelength calibration, sky subtraction and flux calibration. For MUSE data, the archive contains already reduced data. Our data cubes were not collected under photometric conditions, implying our flux calibration is not absolute. However, in order to derive the properties presented in this paper, only a reliable relative flux calibration is required.
\par
Aimed at improving the overall quality of the final datacubes, we performed the data treatment steps described in \citet{Menezes+19}: high spatial-frequency components removal with the Butterworth spatial filtering and ``instrumental fingerprint'' removal via Principal Component Analysis (PCA). After applying the Butterworth filter, we analyzed the residual subtraction between the treated and the untreated datacube. This residual datacube has always an average value very close to zero, implying that photometry is preserved when applying the Butterworth filter.
\par
Also, in order to test the consistency of our results, we performed the same analysis presented in this paper before applying these data treatment procedures. As expected, our main results still hold without them. The figures are just noisier and show stronger vertical bands. These bands are caused by scattered light of the bright AGN, which contaminates the light of the fainter host galaxy. These fingerprints are not fully removed by the reduction script, and thus need additional treatment.
\par
For four objects (NGC\,1068, NGC\,2992, NGC\,3393 and NGC\,5728), we have datacubes from both GMOS and MUSE. We compared the results from these instruments, and found that they closely reproduce each other. In the end, we opted to use the MUSE datacubes solely because of the bigger FoV compared to GMOS. The fact that both datasets agree strengthen our trust on both datasets, since it shows that our results are stable against large perturbations, i.e. changing telescope and instrument.
 \par
Both GMOS and MUSE datacubes are seeing-limited, with spatial point spread function varying between 0\farcs6 and 1\farcs1, measured from point sources in the acquisition image. All GMOS datacubes were obtained using the B600 grating, resulting in a spectral resolution R=$\lambda/\Delta\lambda$ of 1688, and with a 0\farcs1$\times$0\farcs1 spaxel sampling. MUSE data was obtained with the wide field mode, at an average spectral resolution of 1770 at 480\,nm and 3590 at 930\, nm, and with a 0\farcs2$\times$0\farcs2 spaxel sampling. Since the Field of View (FoV) of MUSE (60\farcs0$\times$60\farcs0) is considerably larger than that of GMOS (3\farcs5$\times$5\farcs0), we extracted the inner 5\farcs0$\times$5\farcs0 of MUSE datacubes, centered on the brightest spaxel, in order to properly compare their data for the whole sample.

\setlength{\tabcolsep}{3pt}

\begin{table}%[b]
\centering
%\begin{small}
\caption{The sample and main properties: (1) galaxy name (2) redshift (3) Morphology (4) log of X-ray luminosity in ergs\,s$^{-1}$ (5) source of optical data (6) continuum signal to noise ratio (S/N) for the 10\% lowest flux spaxels (7) continuum S/N for the 10\% highest flux spaxels}
\vspace{0.3cm}
\begin{tabular}{l c H c c c c c}
\hline
Galaxy 	 &$z$        &  Activity   & Morp  &log       & Source & S/N$_{10}$ & S/N$_{90}$\\%   &Gemini Project & MUSE data\\
         &           &             &       &($L_X$)   & of data &            &           \\
\hline
%Mrk1157  &  0.015    &      Sy2   &       &  41.10   &  ---  &  --- &  ---       \\%   & ---                  & ---\\
%NGC788   &  0.014    &      Sy2   &       &  43.20   &  ---  &  --- &  ---       \\%   & ---                  & ---\\
NGC1052   &  0.005    &      Sy2   &  E4   &  41.90   &  GMOS &  21  &  34        \\%   & {\bf GS-2013B-Q-20}  & ---\\
NGC1068   &  0.004    &      Sy2   &  SAb  &  41.80   &  MUSE &  43  &  58        \\%   & {\bf GS-2010B-Q-81}  & Yes\\
%NGC1125  &  0.011    &      Sy2   &       &  42.30   &  ---  &  --- &  ---       \\%   & ---                  & ---\\
Mrk1066   &  0.012    &      Sy2   &  SB0  &  41.1    &  GMOS &  4   &  36        \\%   & {\bf GN-2017B-Q-44}  & ---\\
NGC1194   &  0.013    &      Sy1   &  SA0  &  42.70   &  MUSE &  7   &  17        \\%   & ---                  & Yes\\
Mrk607    &  0.009    &      Sy2   &  Sa   &  40.83   &  GMOS &  8   &  18        \\%   & {\bf GN-2014B-Q-87}  & ---\\
NGC2110   &  0.008    &      Sy2   &  SAB0 &  43.30   &  GMOS &  5   &  18        \\%   & {\bf GN-2017B-Q-44}  & ---\\
%Mrk3     &  0.014    &      Sy2   &       &  43.40   &  ---  &  --- &  ---       \\%   & ---                  & ---\\
%Mrk79    &  0.022    &      Sy1   &       &  43.50   &  ---  &  --- &  ---       \\%   & ---                  & ---\\
NGC2992   &  0.008    &      Sy2   &  Sa   &  42.20   &  MUSE &  12  &  39        \\%   & {\bf GS-2018A-Q-208} & Yes\\
%NGC3035  &  0.015    &      Sy1   &       &  42.70   &  ---  &  --- &  ---       \\%   & ---                  & ---\\
NGC3081   &  0.008    &      Sy2   &  SAB0 &  42.70   &  MUSE &  29  &  38        \\%   & ---                  & Yes\\
%NGC3227  &  0.004    &      Sy1.5 &       &  42.30   &  ---  &  --- &  ---       \\%   & {\bf GN-2013A-Q-61}  & ---\\
NGC3393   &  0.013    &      Sy2   &  SBa &  42.70   &  MUSE &  32  &  45        \\%   & {\bf GS-2015A-Q-12}  & Yes\\
NGC3516   &  0.009    &      Sy1.5 &  SB0  &  43.00   &  GMOS &  12  &  18        \\%   & {\bf GN-2013A-Q-61}  & ---\\
NGC3786   &  0.009    &      Sy1.8 &  SABa &  42.20   &  GMOS &  3   &  14        \\%   & {\bf GN-2013A-Q-61}  & ---\\
%NGC4051  &  0.002    &      Sy1   &       &  41.50   &  ---  &  --- &  ---       \\%   & ---                  & ---\\
%NGC4151  &  0.003    &      Sy1.5 &       &  42.80   &  ---  &  --- &  ---       \\%   & ---                  & ---\\
%NGC4235  &  0.008    &      Sy1   &       &  42.30   &  ---  &  --- &  ---       \\%   & {\bf GS-2018A-Q-232} & ---\\
%Mrk766   &  0.013    &      Sy1.5 &       &  42.80   &  ---  &  --- &  ---       \\%   & ---                  & ---\\
%NGC4388  &  0.008    &      Sy2   &       &  43.30   &  ---  &  --- &  ---       \\%   & ---                  & ---\\
NGC4939   &  0.010    &      Sy1   &  SAbc &  42.40   &  GMOS &  15  &  25        \\%   & {\bf GS-2018A-Q-225} & ---\\
%NGC5506  &  0.006    &      Sy1.9 &       &  43.10   &  ---  &  --- &  ---       \\%   & ---                  & ---\\
%NGC5548  &  0.017    &      Sy1   &       &  43.40   &  ---  &  --- &  ---       \\%   & ---                  & ---\\
NGC5728   &  0.009    &      Sy2   &  SABa &  43.00   &  MUSE &  35  &  53        \\%   & {\bf GS-2013A-Q-56}  & Yes\\
NGC5899   &  0.009    &      Sy2   &  SABc &  42.10   &  GMOS &  8   &  12       \\%   & {\bf GN-2013A-Q-61}  & ---\\ 
%NGC5929  &  0.008    &      Sy2   &       &  41.4    &  ---  &  --- &  ---       \\%   & ---                  & ---\\
\hline	
\\													       
\end{tabular}
 \label{tab:sample}  
\end{table}

\section{Method}
\label{sec:method}
\subsection{Stellar Population}

In order to derive the stellar populations of galaxies, it is generally employed a computing code which fits the underlying continuum with a library of simple stellar population (SSP) models. Many papers have discussed the implications of choosing different codes and libraries \citep[e.g.][]{Chen+10,Ferre-Mateu+12,Maksym+14,Wilkinson+15,Mentz+16}. The different approaches have been shown to yield rather consistent results, as long as the fitting procedure is kept the same. Also, since each code/library may have its own systematic biases, the best way to deal with them is to keep the fitting procedures the same \citep{Goddard+17}.
\par
We employed the {\sc starlight} code \citep{CF+04,CF+05,CF+13}, which fits the observed spectrum by combining stellar population models in different proportions. The code also searches for dust reddening and line of sight velocity distributions that best describe the observed spectrum.
\par
We masked regions dominated by emission lines or with low S/N. Also, in order to deal with weak emission lines not detected in the masking process, we enabled sigma-clipping. This option automatically masks regions in which the observed flux differs from the modeled one by a factor of three sigma.
\par
\begin{figure*}
    %\centering
    \includegraphics[width=\textwidth]{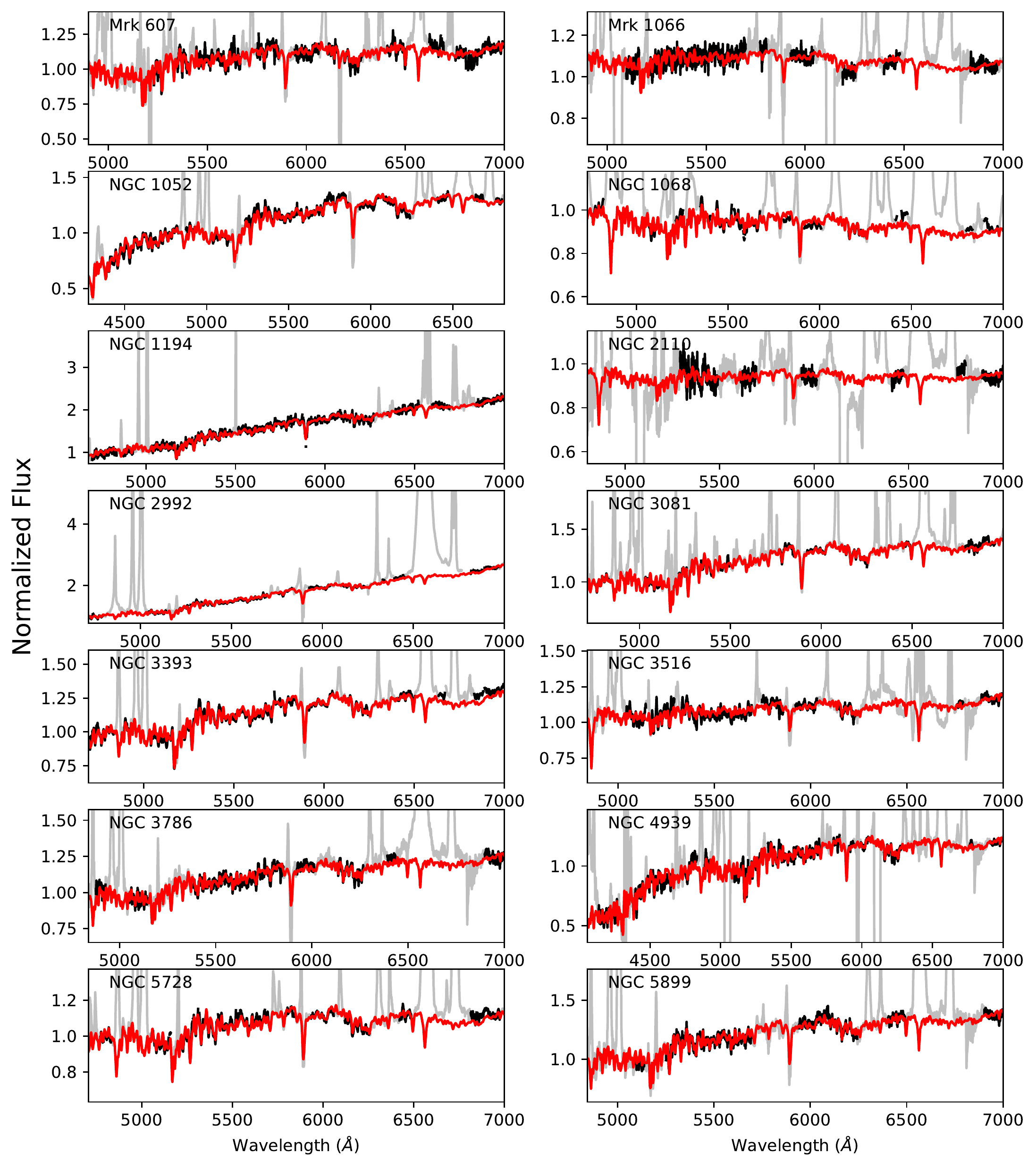}
    \caption{Observed and modeled spectra for the central spaxel of each galaxy in our sample. Black regions represent points used to fit the data, whereas grayed regions represent masked spectra. The spectra modeled by {\sc starlight} are shown in red.}
    \label{fig:fits}
\end{figure*}

We fed {\sc starlight} with the {\sc granada}+MILES (GM) library, well described by \citet{CF+14}. This base was constructed by combining \citet{Vazdekis+10} models, which start at an age of 63\,Myr, with the \citet{GonzalezDelgado+05} models for younger ages. The final set of models consist of stellar templates of 21 ages and 4 metallicities\footnote{Ages t=0.001, 0.00562, 0.01, 0.0141, 0.02, 0.0316, 0.0562, 0.1, 0.2, 0.316, 0.398, 0.501, 0.631, 0.708, 0,794, 0.891, 1.0, 2.0, 5.01, 8.91 and 12.6\,Gyr and metallicities Z=0.2, 0.4, 1 and 1.5\,Z$_\odot$}. In order to reproduce the featureless continuum emitted by the AGN, we first tested a scenario with three power-laws, with $F_\nu \propto \nu^{-1.0}$, $F_\nu \propto \nu^{-1.5}$ and $F_\nu \propto \nu^{-2.0}$. On all cases, the dominant power-law was the intermediate one, and thus we removed the other two and performed the synthesis again. Also, in order to deal with dust reddening, we employed the \citet{Calzetti+00} extinction law. We show in Fig~\ref{fig:fits} the observed and fitted spectra of the central spaxel for each galaxy in our sample. Lastly, due to this different wavelength coverage , we normalized our spectra around 5100\r{A}, with a 100\r{A} window.

\par
Since noise present in the data may cause small variations in the detection of SSPs by {\sc starlight} \citep{CF+04,CF+05}, we present our results in the form of condensed population vectors. Following \citet{Riffel+09}, we defined the light fraction population vectors as: $xy$ (t$\leqslant$0.05\,Gyr), $xi$ (0.05$<$t$\leqslant$2\,Gyr) and $xo$ (t$>$2\,Gyr) to represent the young, intermediate-age and old stellar populations, respectively. We also defined $<Z>_L$ as the luminosity-weighted average metallicity, in units of solar metalicity ($Z_\odot$), and $<t>_L$ as the log of the luminosity-weighted average age, in years.
\par

\subsection{Stellar Kinematics}
With {\sc starlight}, we also mapped stellar centroid velocity (V) and velocity dispersion ($\sigma$) of our sample. In order to check if the V map can be reproduced by a rotating disk, we followed \citet{Bertola+91} and assumed the rotational velocity at a radial distance $r$ from the center of the galaxy to be given by:

\begin{equation} 
v_c = \frac{Ar}{(r^2 + c_0^2)^{p/2}},
\end{equation} 
\noindent
where $A$, $c_0$ and $p$ are parameters to be found. Since we can only measure the velocity component perpendicular to the plane of the sky, the observed radial velocity at a position ($R$,$\Psi$) is then given by:

\scriptsize
\begin{equation}
v(R, \Psi) = v_{sys} + \frac{AR cos(\Psi-\Psi_0)sin\theta cos^p \theta}{\{R^2[sin^2(\Psi-\Psi_0)+cos^2\Theta cos^2(\Psi-\Psi_0)]+c_0^2 cos^2\Theta  \}^{p/2}},
\end{equation}
\normalsize
\noindent
where $v_{sys}$ is the systemic velocity of the galaxy, $\Theta$ is the disk inclination (where $\Theta$=0 represents a face-on disk) and $\Psi_0$ is the position angle of the line of nodes. The fitting of the data was performed using a {\sc python} script employing Levenberg-Marquardt $\chi^2$ minimization.

\section{Results and individual discussion}
\label{sec:results}

\begin{figure*}
    %\centering
    \includegraphics[width=\textwidth]{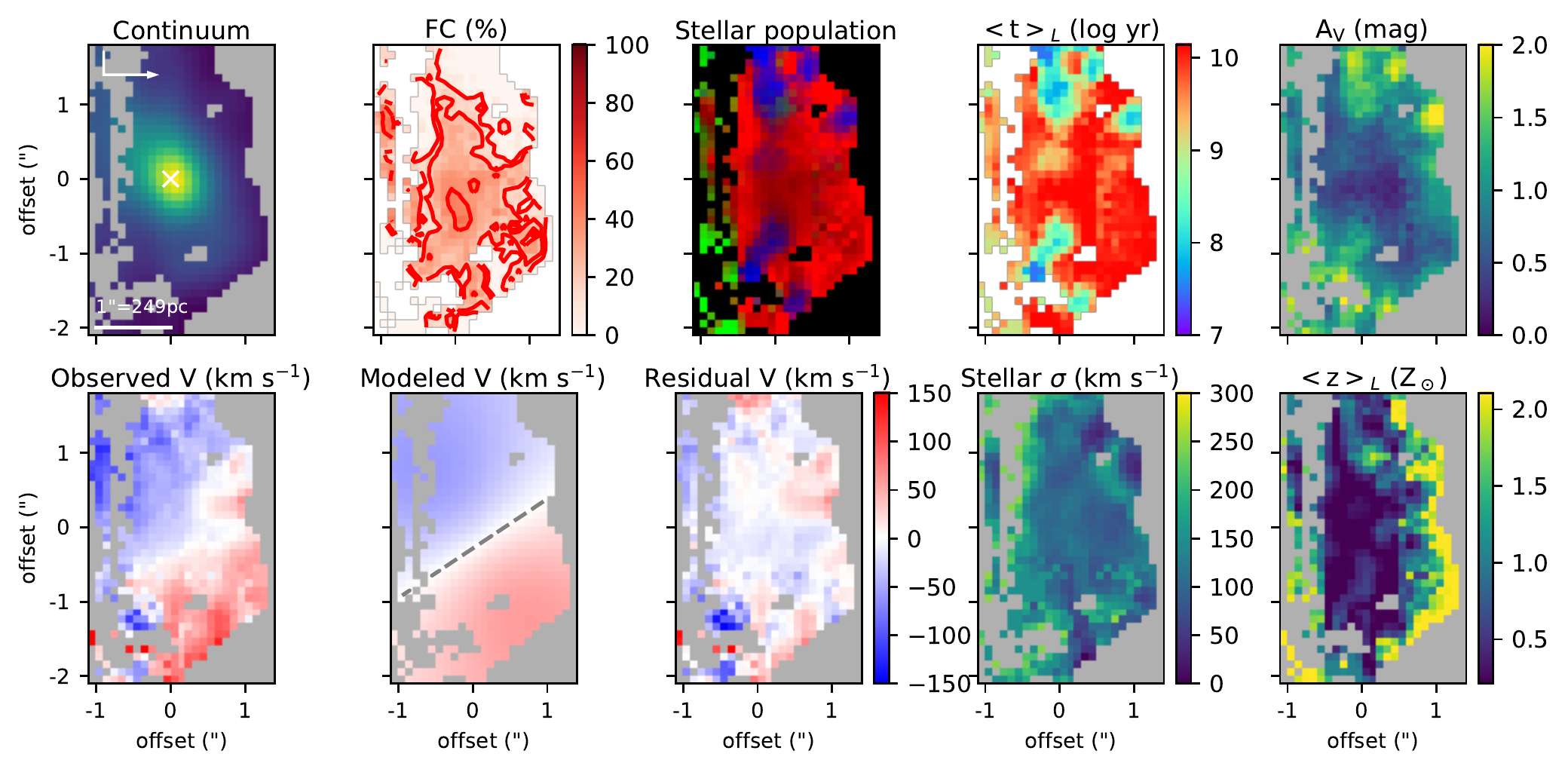}
    \caption{Stellar population results for the SB0 Seyfert\,2 galaxy Mrk\,1066. In the first row, we show the continuum emission in the first left panel. Over this panel the white arrow represents north, whereas east is represented by the small line. In the second panel, we show the percentage contribution from a Featureless continuum, with contours drawn at 10, 20, 40 and 80\%. In the third panel, we show an RGB map composed by the the detected stellar population, where blue represents the young SPs, green the intermediate-age SPs and red the old SPs. The logarithm of the mean age in years is presented in the fourth panel, and the V band dust reddening in magnitudes is presented in the fifth panel. In the second row, we show the observed and modeled stellar velocity (in units of km\,s$^{-1}$) in the first and second panels, respectively, with the residual between the observed and modeled velocities presented in the third panel. Over the modeled V panel, the gray dashed line represents the zero velocity line. In the fourth panel, we show the stellar $\sigma$ in km\,s$^{-1}$, and the metallicity is presented in the last panel in units of solar metallicity.}
    \label{fig:m1066}
\end{figure*}

\begin{figure*}
    %\centering
    \includegraphics[width=\textwidth]{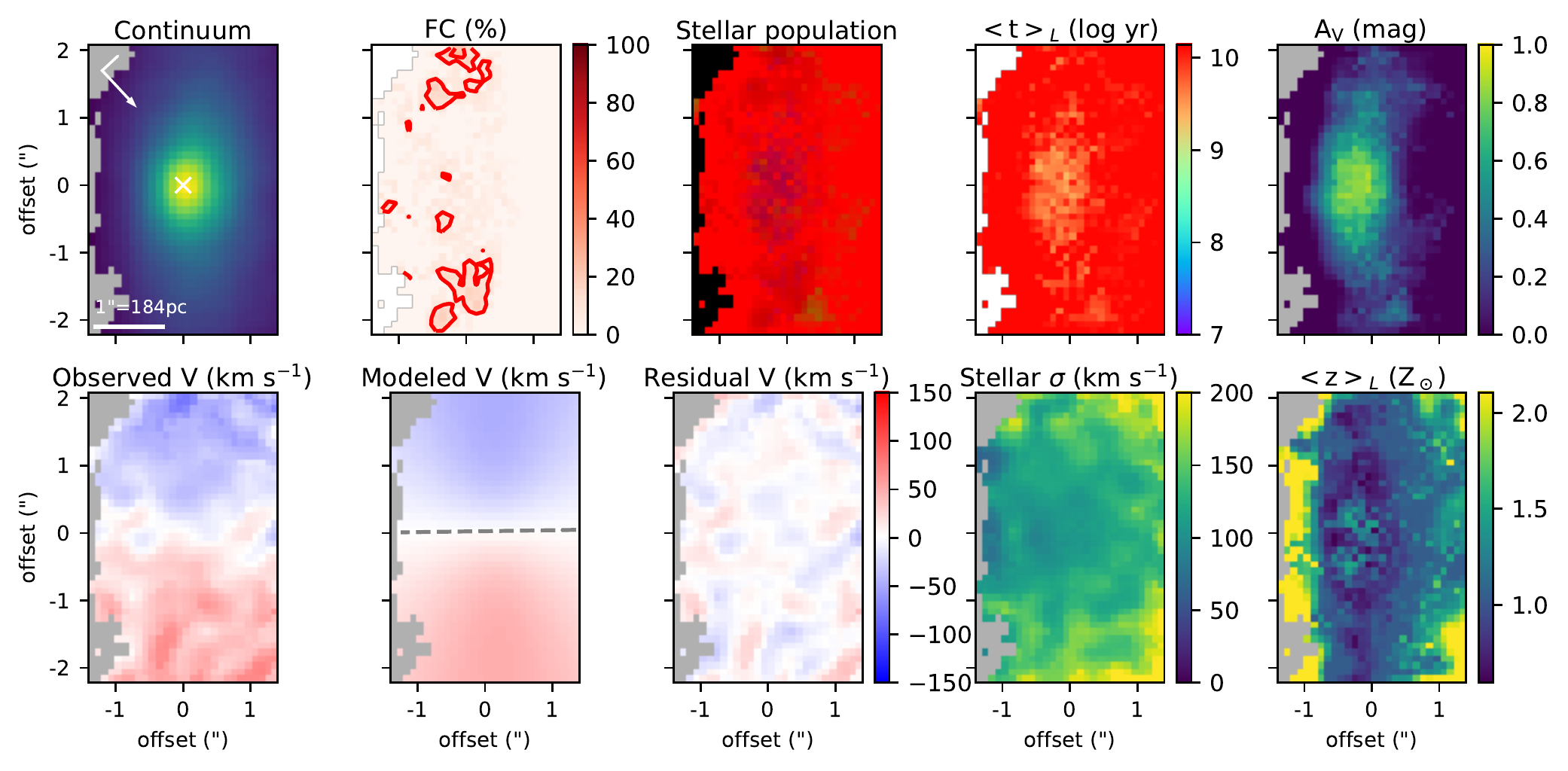}
    \caption{Same as Fig~\ref{fig:m1066}, but for Sa Seyfert\,2 galaxy Mrk\,607.}
    \label{fig:m607}
\end{figure*}

\begin{figure*}
    %\centering
    \includegraphics[width=\textwidth]{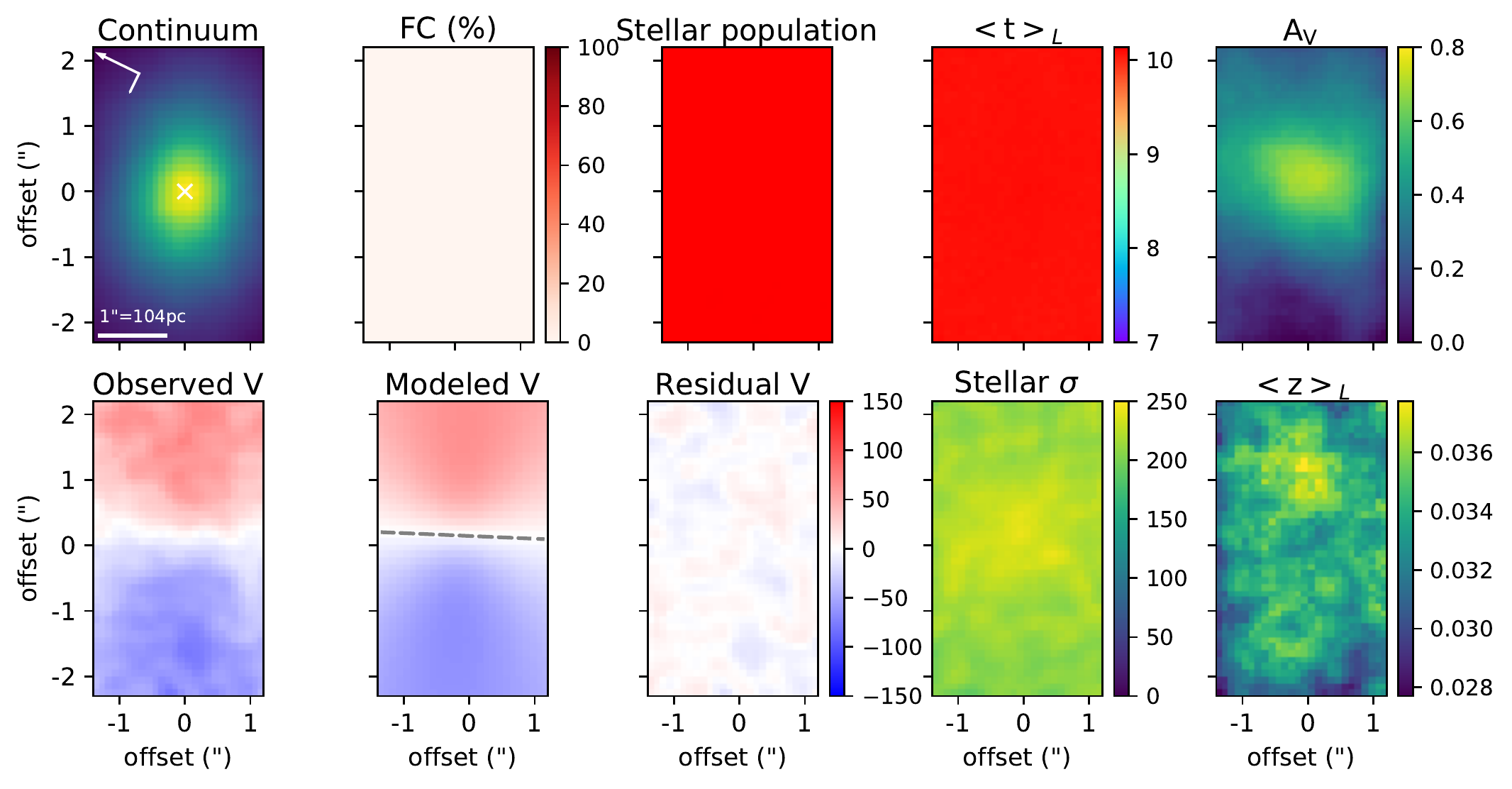}
    \caption{Same as Fig~\ref{fig:m1066}, but for E4 LINER/Seyfert galaxy NGC\,1052.}
    \label{fig:n1052}
\end{figure*}

\begin{figure*}
    %\centering
    \includegraphics[width=\textwidth]{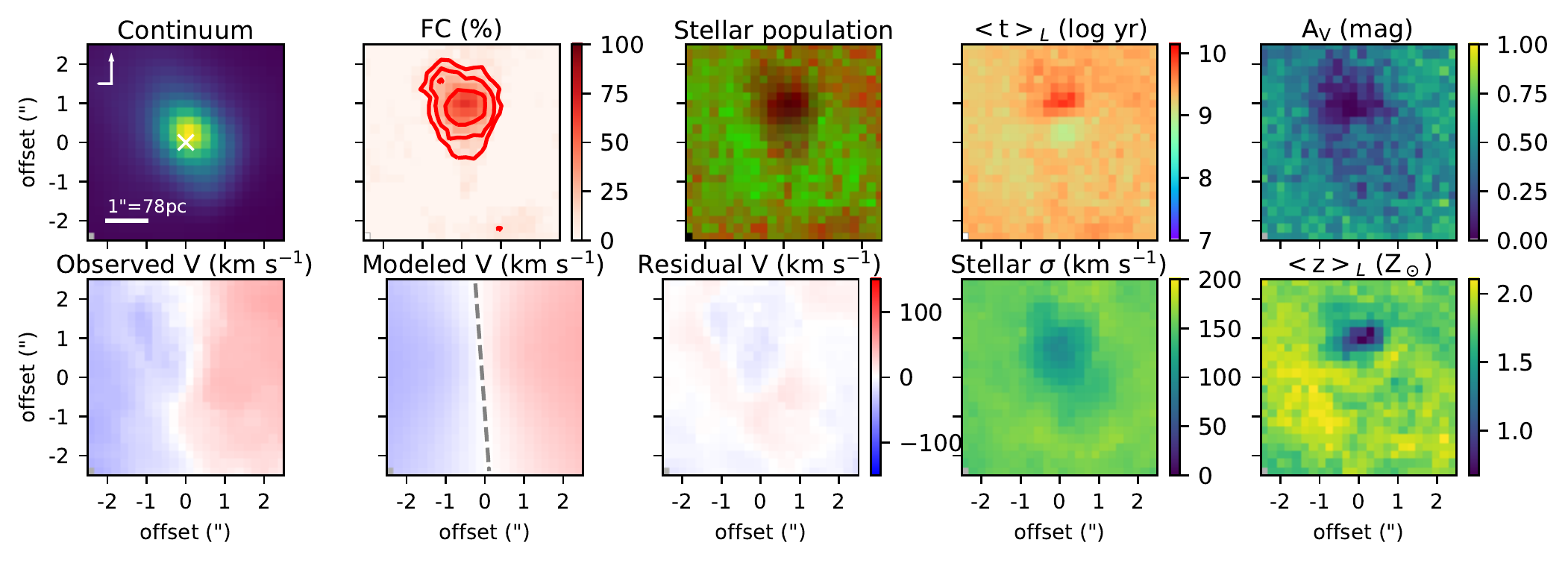}
    \caption{Same as Fig~\ref{fig:m1066}, but for SAb Seyefrt\,2 galaxy NGC\,1068.}
    \label{fig:n1068}
\end{figure*}

\begin{figure*}
    %\centering
    \includegraphics[width=\textwidth]{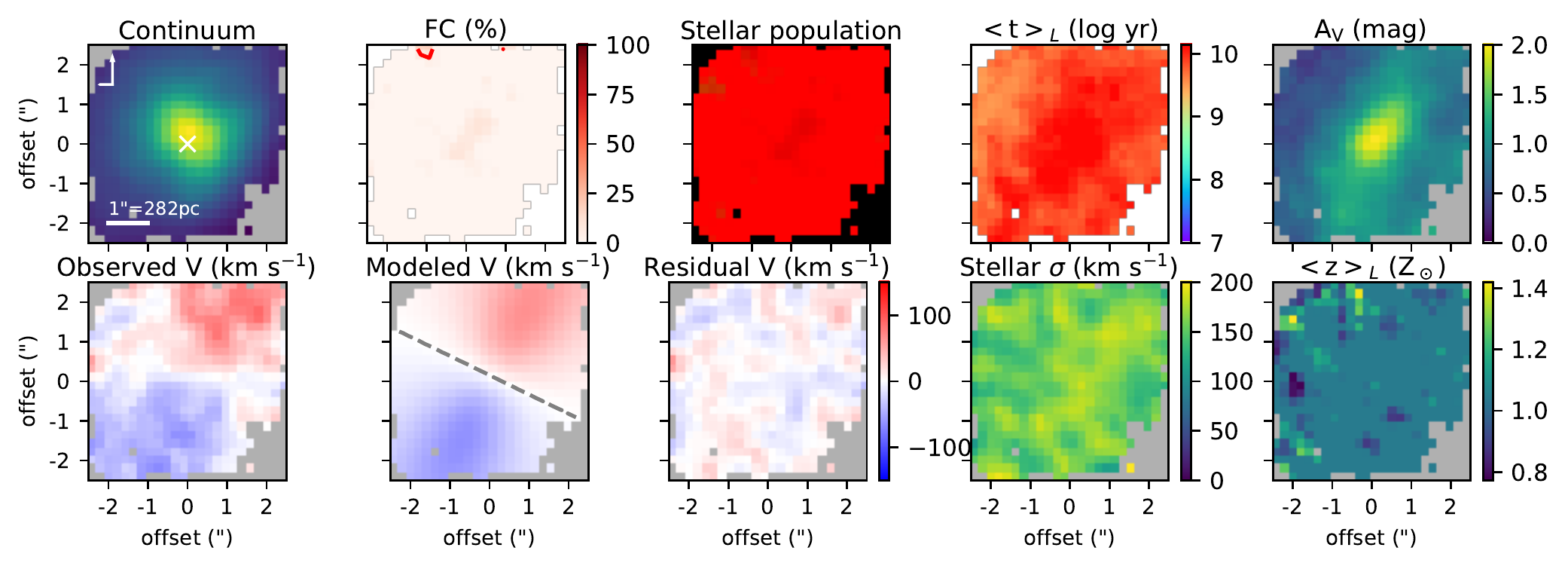}
    \caption{Same as Fig~\ref{fig:m1066}, but for SA Seyfert\,1 galaxy NGC\,1194.}
    \label{fig:n1194}
\end{figure*}

\begin{figure*}
    %\centering
    \includegraphics[width=\textwidth]{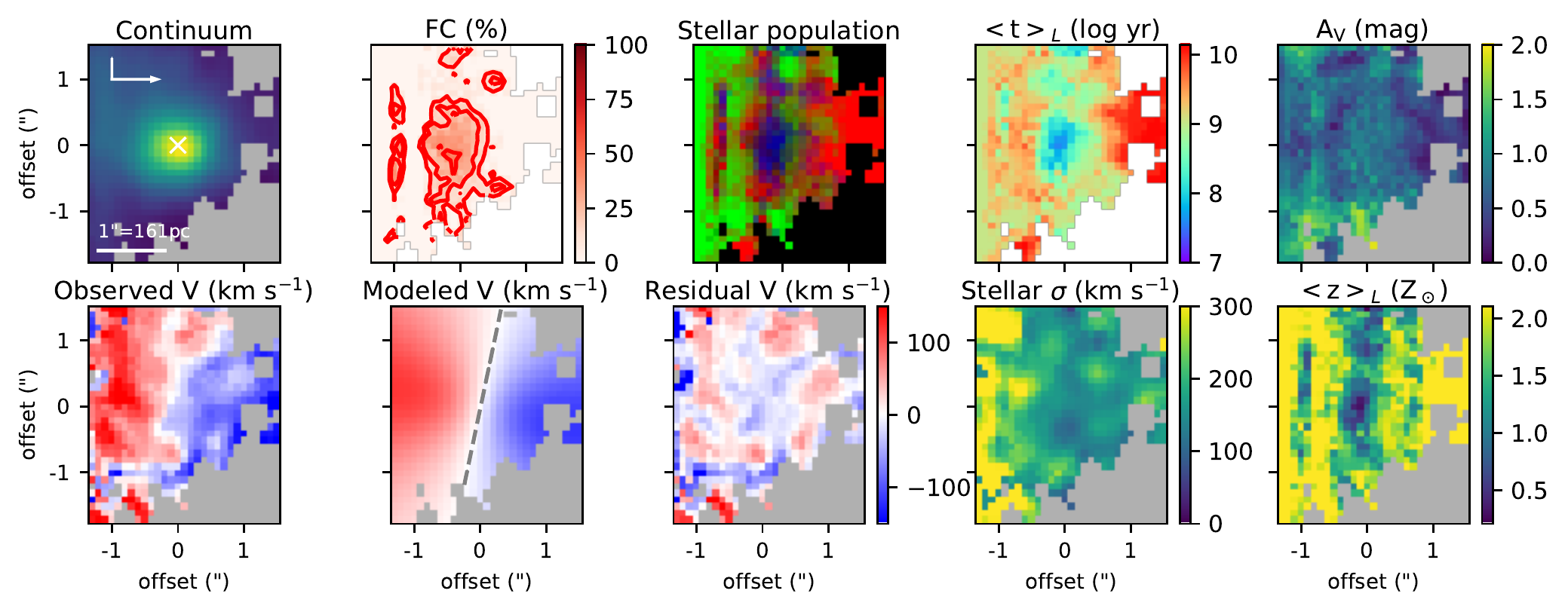}
    \caption{Same as Fig~\ref{fig:m1066}, but for S0 Seyfert\,2 galaxy NGC\,2110.}
    \label{fig:n2110}
\end{figure*}

\begin{figure*}
    %\centering
    \includegraphics[width=\textwidth]{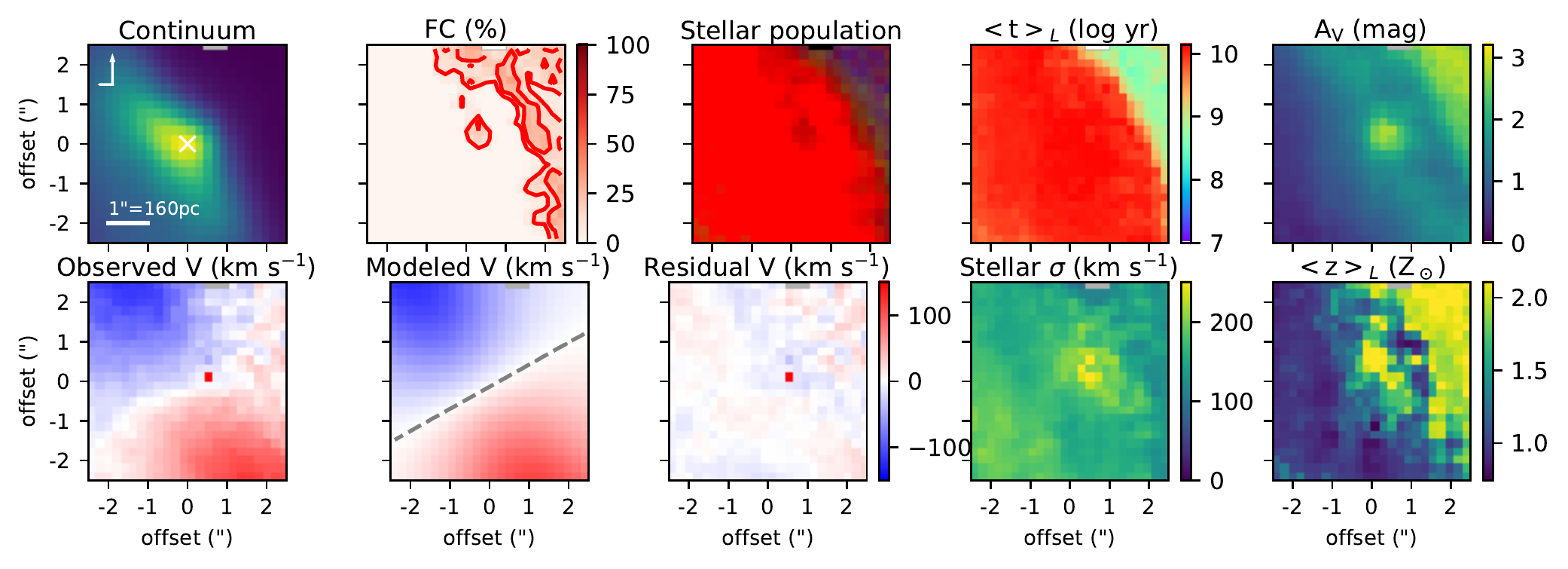}
    \caption{Same as Fig~\ref{fig:m1066}, but for Sa Seyfert\,2 galaxy NGC\,2992.}
    \label{fig:n2992}
\end{figure*}

\begin{figure*}
    %\centering
    \includegraphics[width=\textwidth]{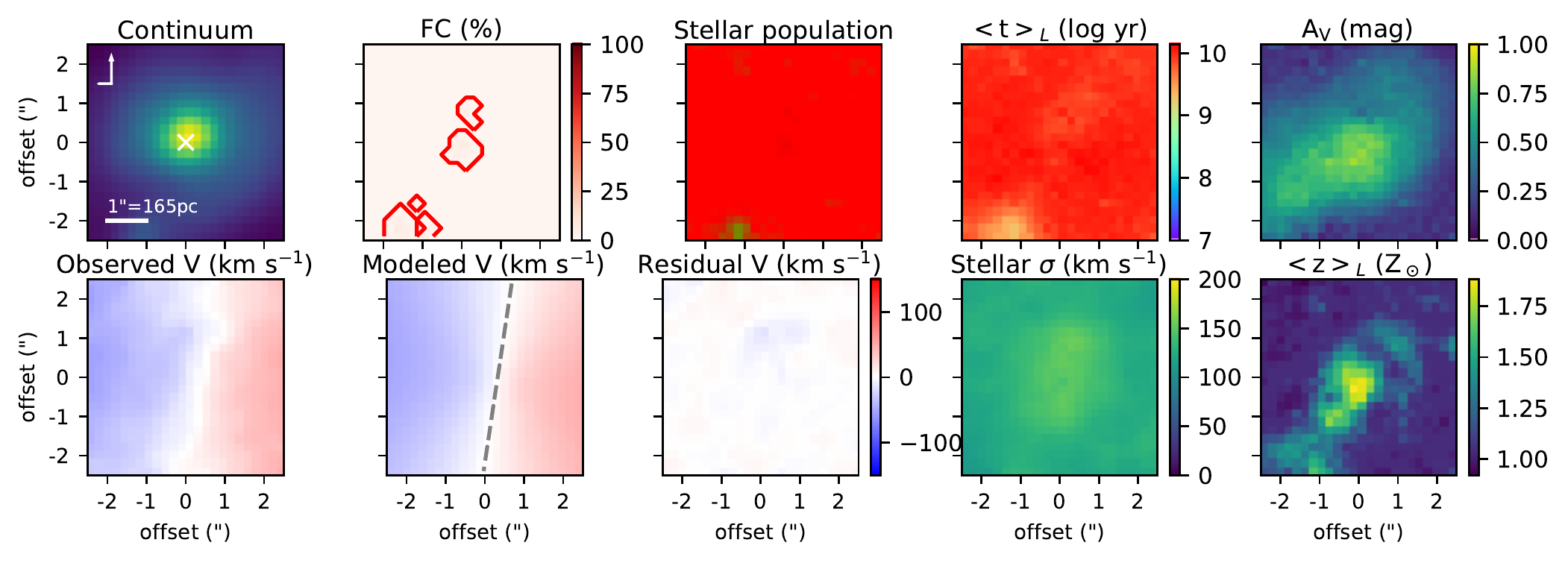}
    \caption{Same as Fig~\ref{fig:m1066}, but for S0 Seyfert\,2 galaxy NGC\,3081.}
    \label{fig:n3081}
\end{figure*}

\begin{figure*}
    %\centering
    \includegraphics[width=\textwidth]{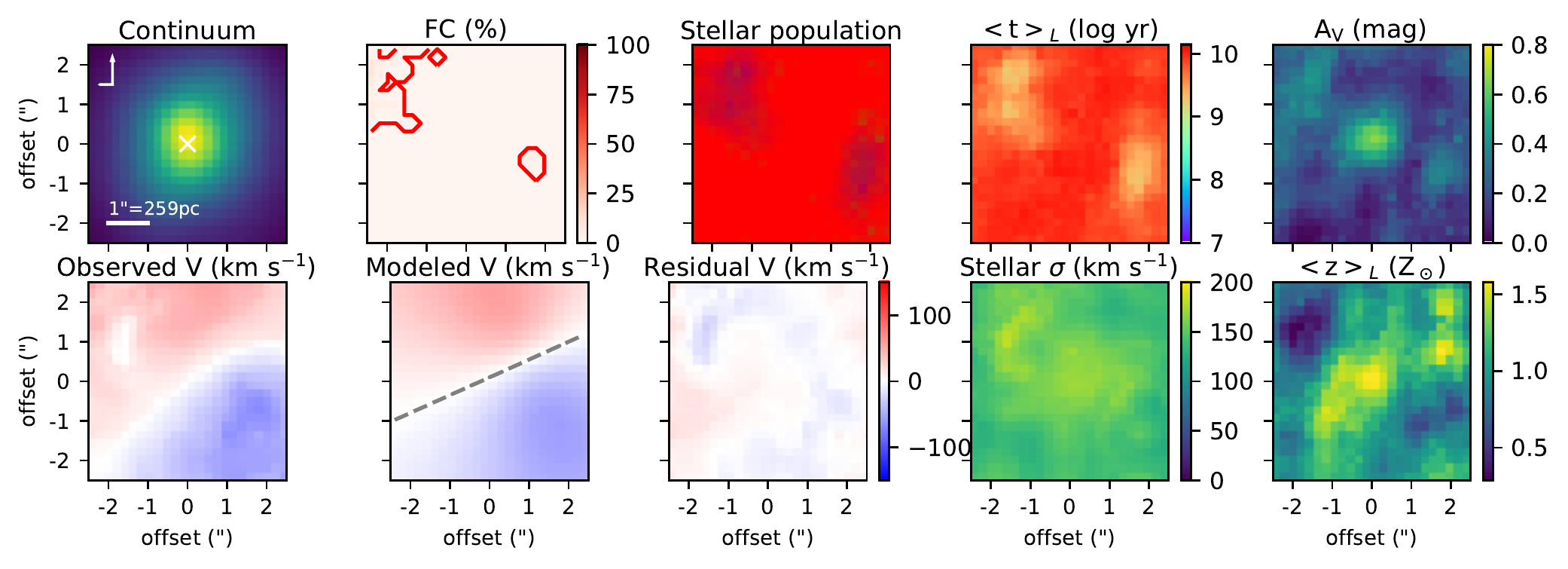}
    \caption{Same as Fig~\ref{fig:m1066}, but for Sa Seyfert\,2 galaxy NGC\,3393.}
    \label{fig:n3393}
\end{figure*}

\begin{figure*}
    %\centering
    \includegraphics[width=\textwidth]{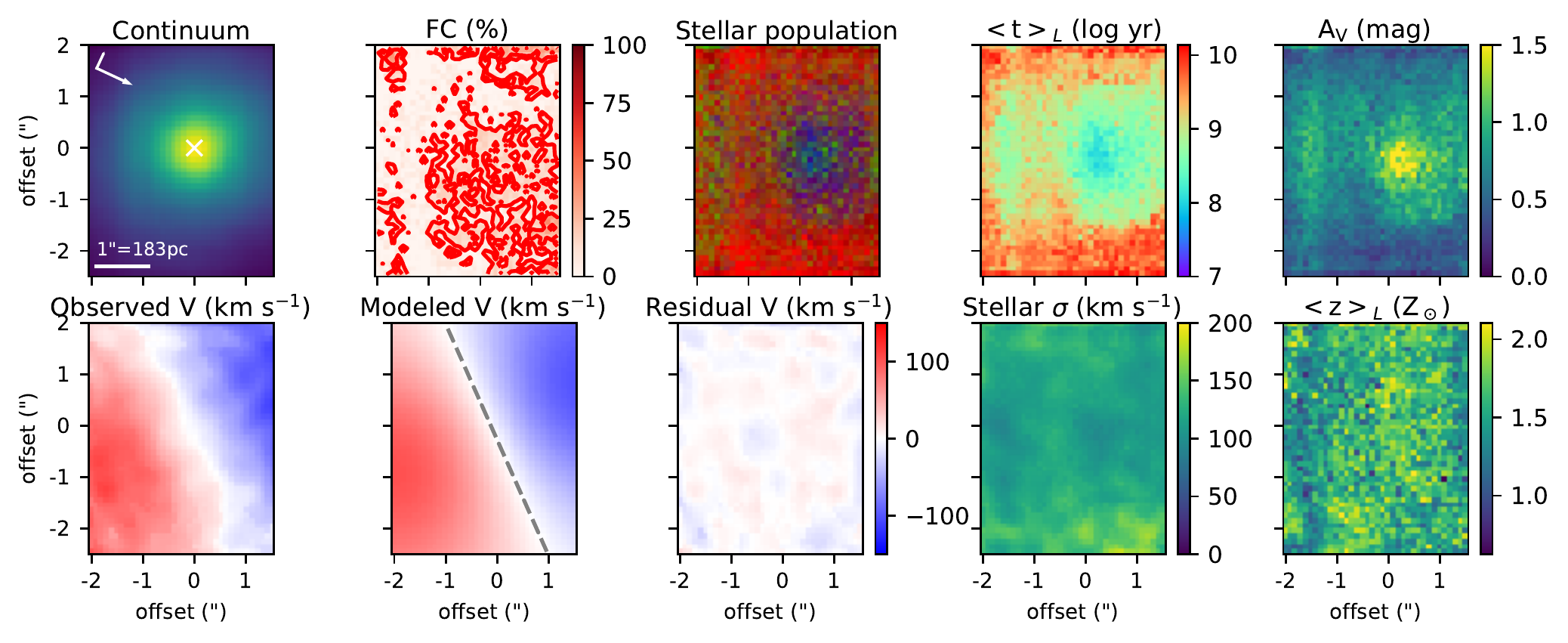}
    \caption{Same as Fig~\ref{fig:m1066}, but for SB0 Seyfert\,1.5 galaxy NGC\,3516.}
    \label{fig:n3516}
\end{figure*}

\begin{figure*}
    %\centering
    \includegraphics[width=\textwidth]{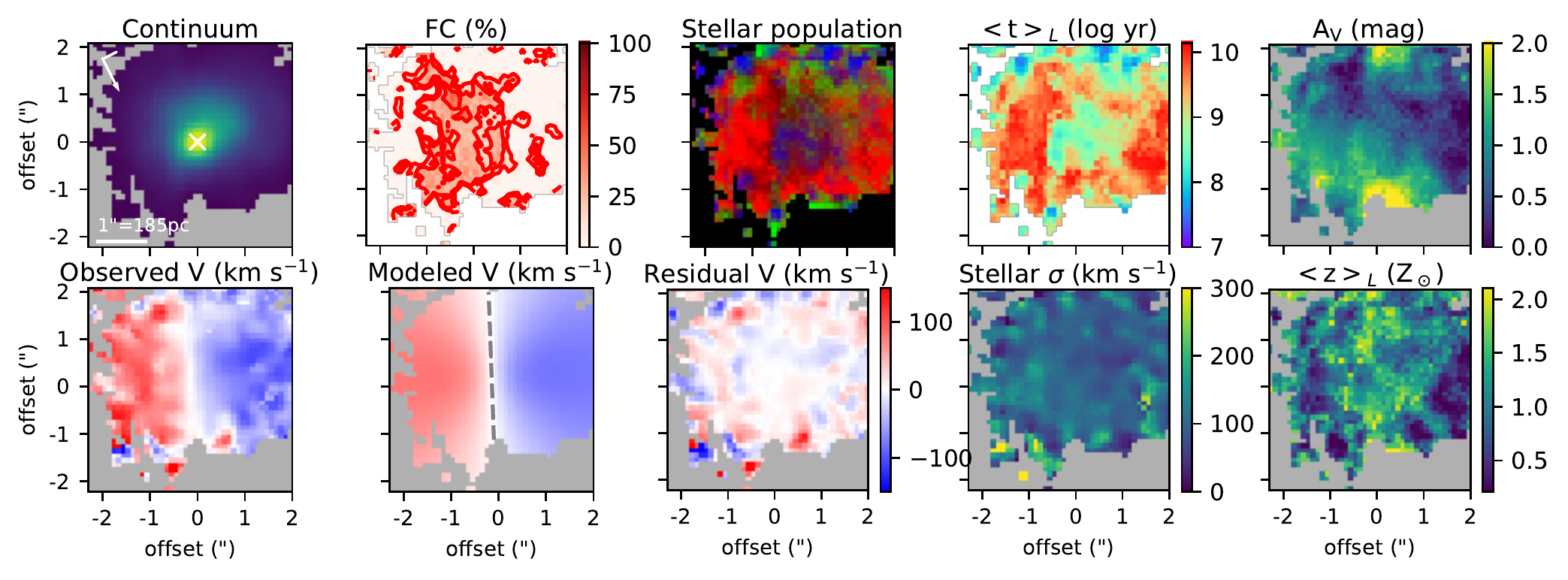}
    \caption{Same as Fig~\ref{fig:m1066}, but for SA Seyfert\,1.8 galaxy NGC\,3786.}
    \label{fig:n3786}
\end{figure*}

\begin{figure*}
    %\centering
    \includegraphics[width=\textwidth]{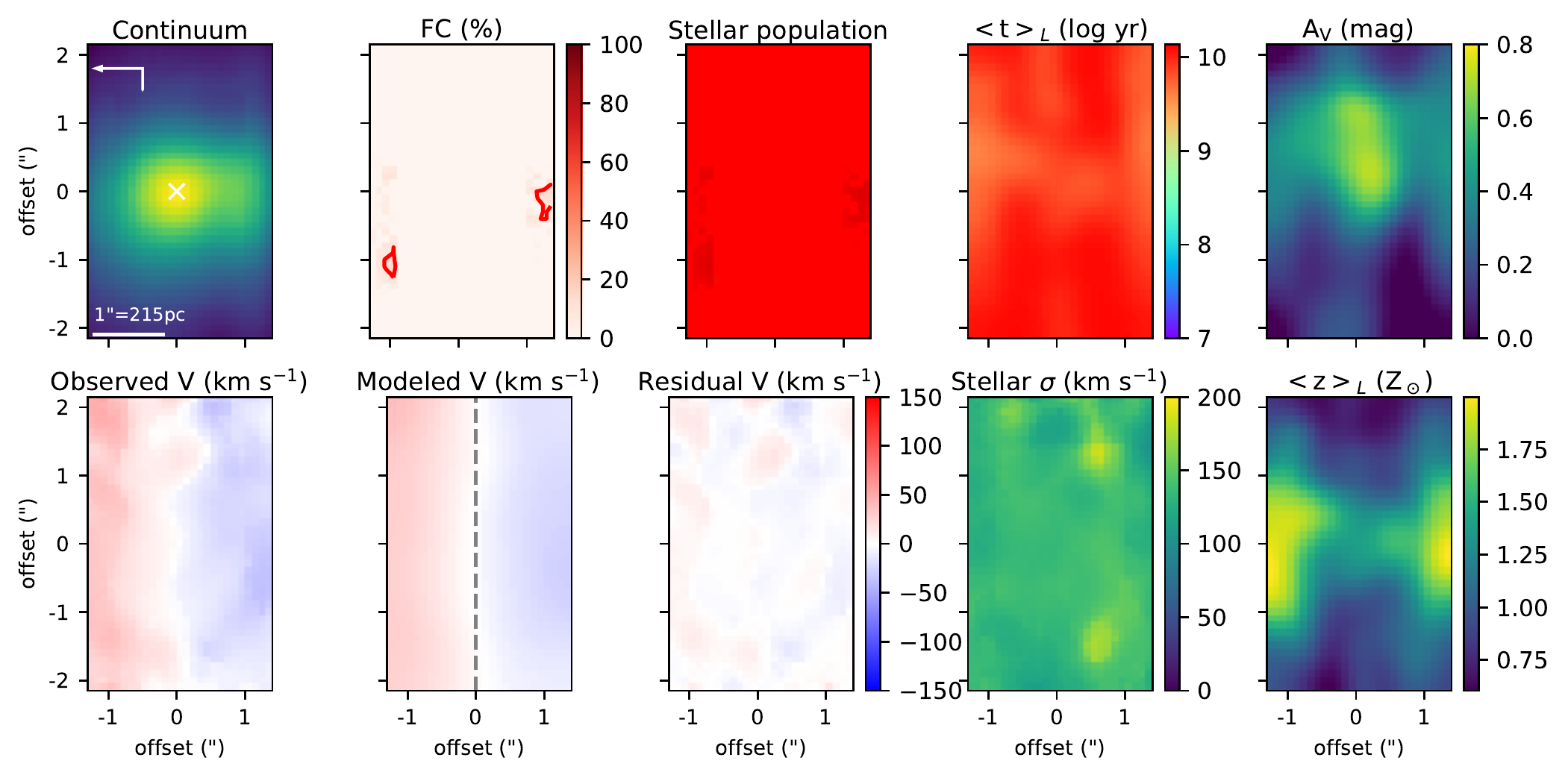}
    \caption{Same as Fig~\ref{fig:m1066}, but for Sbc Seyfert\,1 galaxy NGC\,4939.}
    \label{fig:n4939}
\end{figure*}

\begin{figure*}
    %\centering
    \includegraphics[width=\textwidth]{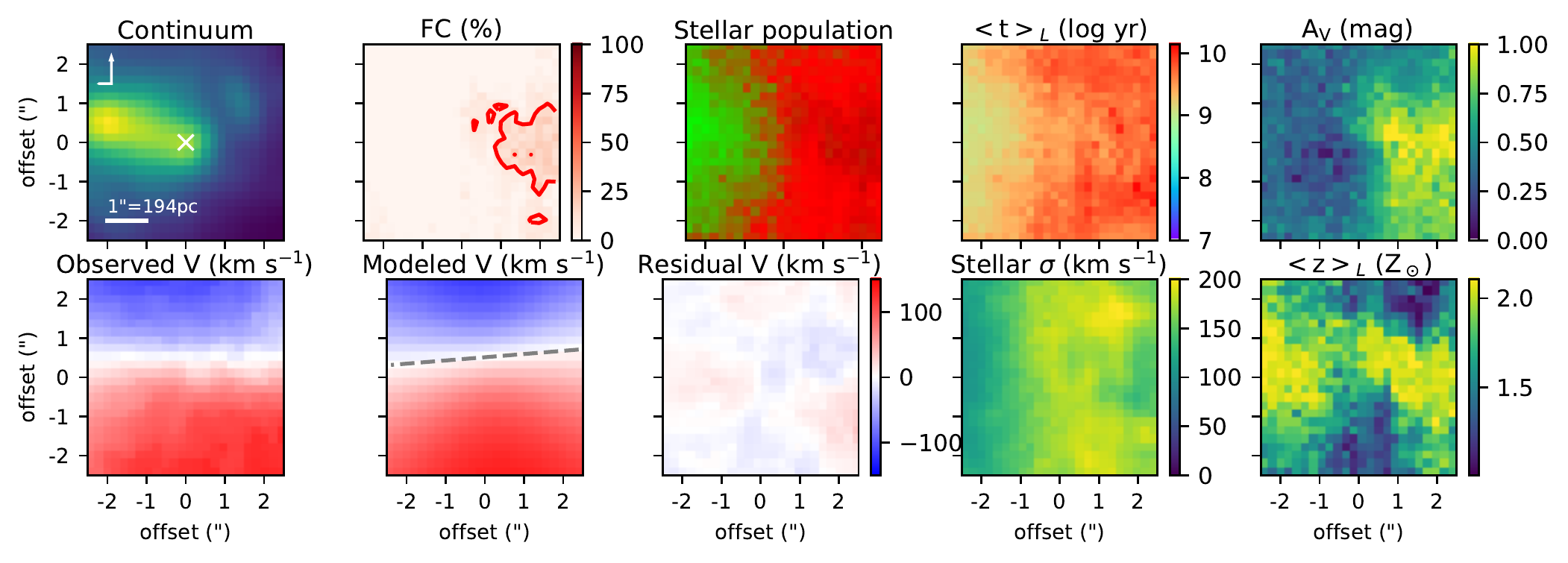}
    \caption{Same as Fig~\ref{fig:m1066}, but for SBb Seyfert\,2 galaxy NGC\,5728.}
    \label{fig:n5728}
\end{figure*}

\begin{figure*}
    %\centering
    \includegraphics[width=\textwidth]{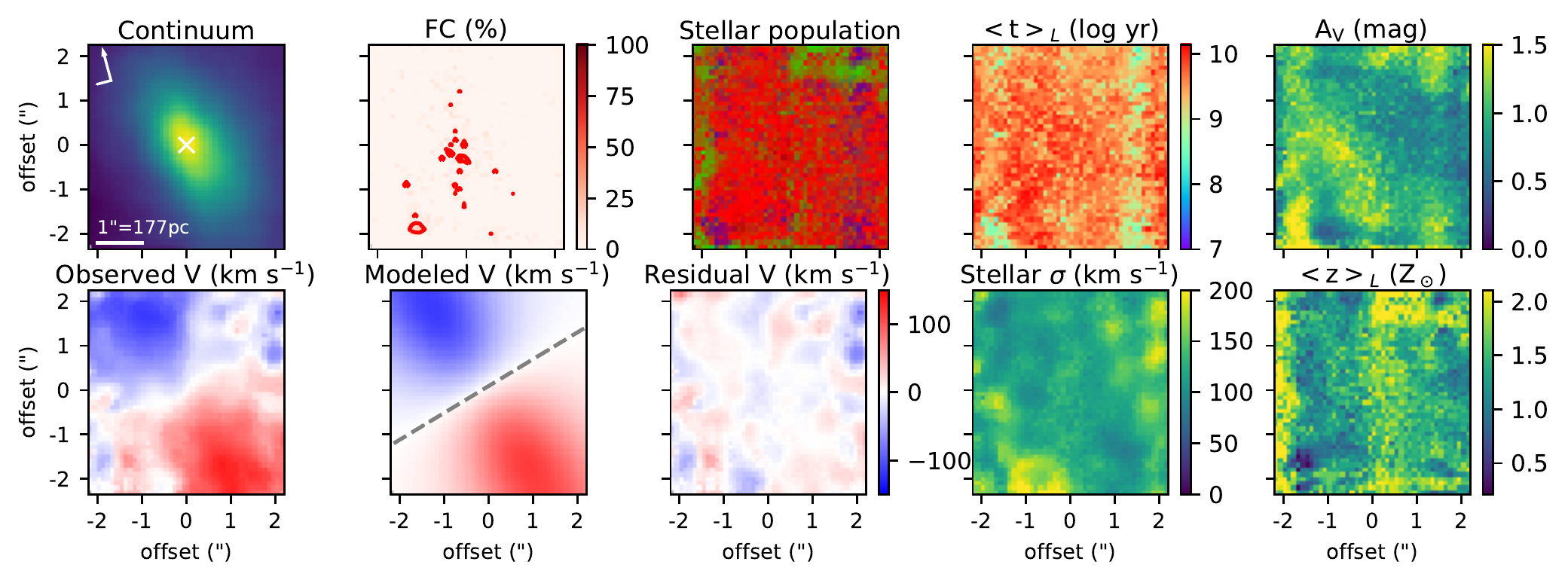}
    \caption{Same as Fig~\ref{fig:m1066}, but for SAB Seyfert\,2 galaxy NGC\,5899.}
    \label{fig:n5899}
\end{figure*}

The individual results of our sample are presented in Figs~\ref{fig:m1066} to \ref{fig:n5899} with the following structure: in the top row, we present in consecutive panels, from left to right, the integrated continuum emission within the full spectral window used by {\sc starlight}, the percentage contribution of the featureless continuum (FC) in the second panel; in the third panel, we show RGB maps where red, green and blue represent xo, xi and xy respectively; in the fourth and fifth panels, we show $<t>_L$ and the dust extinction in band V, respectively; in the bottom row we show in the first and second panels the measured and modeled stellar velocities, whereas in the third panel we show the residual of their subtraction; in the fourth panel we show the stellar $\sigma$, and in the fifth panel the $<z>_L$.
\par
Since our data comes from different instruments and different projects, with also diverse objectives, the S/N of our sample varies greatly, both in the emission lines and the continuum level. We then decided to mask regions with S/N<6. This value was determined by analyzing at which point the fits became unreliable. Also, due to the low S/N in extranuclear regions of part of our sample, we were unable to derive the $h3$ and $h4$ moments of the Gauss-Hermite polynomials, which measure asymmetries in the absorption bands profiles. For this reason, we did not include such results in this study. Lastly, we removed NGC\,3227 and NGC\,4235 from our analysis, the first for being dominated by broad emission lines which precluded a proper fit of the continuum, and the second due to low-order spurious features redward of 5100\,\r{A}. However, we were able to measure the main emission lines for both galaxies, which will be presented in a forthcoming paper (Paper II).
\par
The morphological classifications were taken from \citet{deVaucouleurs+91}, and the nuclear activity classification was taken from \citet{Oh+18}. The nuclear activity classification will be reviewed in Paper II based on our emission-line spectra.

\subsection{Mrk\,1066}
%The GMOS datacube was never analyzed 
This galaxy is classified as SB0 and hosts a Seyfert\,2 AGN. It is one of the galaxies with the lowest S/N of our sample. Nonetheless, it is still possible to detect a very prominent contribution of a FC at the nucleus, with mostly old stellar populations throughout the datacube. Regarding the stellar kinematics, we found that the stars are well reproduced by a disk in rotation, with a slight S-shaped assymetry, and a nearly constant stellar $\sigma$. Lastly, we detected a drop in both A$_{\rm V}$ and metallicity in the nucleus. The fact that we detected a dust extinction drop in the nucleus of this and other galaxies means that the central regions are so heavily obscured that the light we are observing is mainly composed of unextinguished stellar emission in the leading edge of the host galaxy. Also, the stellar kinematics of this galaxy have been analyzed by \citet{RiffelRA+11}, who had also detected an S-shaped distortion, which they attributed to an oval structure in the continuum. %\citet{RiffelRA+10} studied the stellar populations from NIFS J and K datacubes, and detected a mixture of old and young stellar populations in the inner $\sim$1\farcs0, with a significant contribution from a FC, surrounded by a circumnuclear ring with ages between 0.3 and 0.7\,Gyr. The fact that we did not detect this ring might be attributed to dust obscuration, in which case this ring would be heavily affected by dust reddening.

\subsection{Mrk\,607}
Mrk\,607 is an Sa galaxy harboring a Seyfert\,2 nucleus. We detected a very weak ($<$10\%) FC contribution in the nucleus, which slightly rises to the upper and lower regions of the datacube. This result could be caused by the jet encountering higher densities (see section~\ref{sec:1068} for a broader discussion), but given that this contribution is very small, could be as well caused by statistical uncertainties. The stellar population is dominated by older components in the circumnuclear region, with a combination of old and young stellar populations in the inner $\sim$1\farcs0. The dust reddening peaks at $\sim$1\,mag in the nucleus, and decreases to the edges of the datacube. The stellar kinematics are characterized by blueshifts to the southeast and redshifts to the northwest, as well as a lower velocity dispersion in the center if compared to the borders of the datacube. These kinematical properties agree well with the ones derived by \citet{Freitas+18}, who also analyzed this GMOS datacube and reported that the stellar kinematics within our FoV is well reproduced by a disk in rotation with the velocity dispersion rising away from the nucleus. Lastly, the metallicity is also higher to the edges of our FoV.

\subsection{NGC\,1052}
%This datacube has already been analyzed by \citet{luisgdh+19a,luisgdh+19b}, who mapped the stellar and gas properties both in the optical and NIR.
NGG\,1052 is a giant E4 galaxy \citep{Forbes+01,Xilouris+04}, with an ambiguous LINER/Seyfert classification, well known for having one of the nearest radio-loud AGN \citep{Heckman80,Ho+97,RiffelRA+17}.  The fact that it is the only elliptical galaxy in our sample is consistent with the fact that it is the only galaxy with no traces of young and intermediate-age SSPs. Also, the stellar kinematics are also consistent with its elliptical nature, with the highest stellar $\sigma$ ($\sim$250\,km\,s$^{-1}$) and smallest peak velocities ($\sim$60\,km\,s$^{-1}$). The metallicity is uneven, with a peak located 1\farcs0 to the northwest of the nucleus. Also, it is the only object in which no contribution of a FC was detected, which can be taken as a consequence of its dual LINER/Seyfert classification, implying that it harbors a less active AGN if compared to the rest of the sample. Both our stellar populations and kinematics match well the ones derived in the optical by \citet{luisgdh+18}. However, they also mapped the equivalent widths of the stellar absorptions, and were able to detect a very weak drop in these measurements close to the nucleus. This drop was attributed by the authors to a featureless continuum emitted by the AGN.

\subsection{NGC\,1068}
\label{sec:1068}
%The MUSE datacube we used was analyzed by \citet{Mingozzi+19}, but their analysis focused on the emission line properties.
This galaxy is the brightest and thus most studied Seyfert galaxy, located at a distance of only 14.4\,Mpc \citep{Crenshaw&Kraemer00}. {It is classified as Seyfert\,1.9}, with an SAb Hubble classification. Despite being considered the archetypal Seyfert\,2 galaxy, its properties are far from standard. We detected a significant contribution of a featureless continuum peaking 1\farcs0 north of the nucleus. A similar result has already been reported by \citet{Martins+10}, who analysed a NIR long-slit spectrum along the north-south direction passing through the nucleus. They reported finding two peaks in the FC distribution, one to the north and one to the south, co-spatial with peaks of hot dust emission. The authors attributed this offset to a region where the nuclear jet encounters dense clouds, and thus the light emitted by the AGN is scattered by these clouds. The fact that we did not detect the southern peak suggests that it originates in the far side of these walls, and thus this emission is obscured by dust. In this same region, we also found differences in the dust reddening, stellar $\sigma$ and metallicity. However, these differences must be taken with caution, since they might be caused by spectral degeneracies \citep[see][]{Worthey94,CF+05}. Regarding the stellar population, \citet{Martins+10} reported a dominance of intermediate-age stellar populations with an average age between 1 and 2\,Gyr in the region they analyzed.  \citet{StorchiBergmann+12} studied a NIR datacube with a similar FoV as ours, and found that the inner $\sim$0\farcs5 is dominated by stellar populations older than 5\,Gyr, with a circumnuclear ring dominated by stars younger than 0.7\,Gyr. We did not detect any of the features reported by \citet{StorchiBergmann+12}, with the bulk of the emission being dominated by intermediate-aged stellar populations between 1 and 2\,Gyr, similar to the results reported by \citet{Martins+10}. Lastly, the peak velocity of this galaxy is well explained by a disk in rotation.

\subsection{NGC\,1194}
%The MUSE datacube was never analyzed 
NGC\,1194 is an SA0 galaxy, with a Seyfert\,2 AGN. We detected a very weak ($<$10\%) contribution of a FC in the inner $\sim$1\farcs0 region, with the whole datacube dominated by an old stellar population. The outer regions are younger, with a mean age of $\sim$4\,Gyr, if compared to the inner $\sim$2\farcs0 diameter region, that has a mean age of $\sim$10\,Gyr. The reddening is also higher in the nucleus, peaking at 2\,mag, and the stellar $\sigma$ and metallicity are roughly constant within our data ($\sim$180\,km\,s$^{-1}$ and solar metallicity).

\subsection{NGC\,2110}
%The GMOS datacube was never analyzed 
NGC\,2110 is classified as SAB0, harboring a Seyfert\,2 nucleus. In this galaxy, we detected a significant FC contribution of up to 40\% in the inner $\sim$0\farcs6 region. This region is also dominated by a young stellar population, with a $<t>_L$ of 0.1\,Gyr and subsolar metallicity. According to our analysis, both the reddening and the stellar $\sigma$ are constant within our FoV, except from border effects. A similar analysis was performed by \citet{Diniz+19}, based on J and K datacubes. According to their analysis, the inner $\sim$0\farcs3 are dominated by a combination of young stellar populations, a featureless continuum and a blackbody, which is consistent with our analysis. Also, they detected two circumnuclear rings with different stellar populations, one smaller ($\sim$1\farcs0), dominated by old stellar populations, and one extending from the end of this ring up to the edge of their FoV, dominated by stellar populations with ages between 0.1 and 0.7\,Gyr. We did not detect such rings, and the circumnuclear stellar populations within our FoV are mainly dominated by a mixture of intermediate-age and old SSPs.

\subsection{NGC\,2992}
%The MUSE datacube was also analyzed by \citet{Mingozzi+19}, again focusing on the emission line properties. 
This galaxy receives the Sa Hubble classification, and hosts a Seyfert\,1.9 nucleus. In this object, we detected two regions with vastly different properties. One, to the northwest of our FoV, with a $\sim$20\% contribution of a FC, dominated by a young, metal-rich and very reddened stellar population. The other, containing the rest of our FoV, is dominated by an old, metal-poor and less reddened stellar population, with a very feeble contribution from a FC and a stronger dust reddening in the unresolved nucleus. The first region, younger and more reddened, is associated with a prominent dust lane observed in this object \citep{Colina+87}, whereas the older region is associated with the bulge of the galaxy. Also, the strong FC that we detected off-nucleus, co-spatial with an enhanced dust reddening, is consistent with the FC from the AGN being deflected by denser clouds. Also, our results suggest that the stellar kinematics are dominated by circular rotation around the nucleus, with an S-shaped asymmetry.

\subsection{NGC\,3081}
NGC\,3081 is classified as SAB0, with a Seyfert\,2 nucleus. We found that the nucleus displays slightly different properties than the rest of the galaxy, exhibiting a higher reddening ($\sim$0.8\,mag), a higher stellar sigma ($\sim$180\,km\,s$^{-1}$), a higher metallicity ($\sim$0.035\%), and a very faint ($\sim$10\%) contribution of a FC. Regarding the stellar population, we detected a dominance of old SPs, with a very small contribution of intermediate-age SPs in the south-eastern edge of our FoV, co-spatial with a $\sim$10\% contribution of a FC. Very old stellar populations dominate the observed FoV, but not as old as NGC\,1052. These SP results are consistent with the typical central populations of S0 galaxies. The star formation has already been analyzed by \citet{ErrozFerrer+19}, who mapped the SFR of this object through its emission lines, although covering a larger FoV than ours. They found a very large ring with a radius of 3\,kpc, with SFR surface density up to ${10^{-2} {\rm M_\odot yr^{-1} kpc^{-2}}}$. However, since the nuclear region was dominated by AGN emission, they could not determine the SFR within our FoV.

\subsection{NGC\,3393}
This galaxy is classified as SBa, harboring a Seyfert\,2 nucleus. According to our analysis, this object is dominated by old SPs, with two very distinct spiral arms composed of a combination of old and young SPs. In the same region of these arms, a slight contribution from a FC ($\sim$10\%) was detected, which may be associated with light emitted by the AGN and scattered by the arms, or may be actually due to the contribution from very young SPs, whose spectrum is similar to that from an FC. Also, the inner $\sim$1\farcs5 has a higher dust reddening ($\sim$1.2\,mag), stellar $\sigma$ ($\sim$180\,km\,s$^{-1}$) metallicity (${<Z>\sim0.03}$), if compared to the rest of the galaxy. The star formation of this galaxy has already been analyzed by \citet{ErrozFerrer+19}, and similarly to NGC\,3081 they could not map the SFR of this galaxy close to the nucleus, due to the dominant contribution of emission lines from the AGN. In the external regions, they detected very few H{\sc ii} regions. 

\subsection{NGC\,3516}
%GMOS datacube never analyzed
NGC\,3516 is morphologically classified as SB0, hosting a a Seyfert\,1.2 AGN. We detected a significant contribution from a FC (up to 30\%) in the inner 2\farcs0. This region is also dominated by a younger ($<t>_L\sim$8.5) and more reddened SP (A$_{\rm V}\sim$1.5\,mag), if compared to the rest of our FoV ($<t>_L\sim$9.5 and A$_{\rm V}\sim$0.5\,mag). We also found that the kinematics of the stars are well described by a disk in rotation, with a constant ($\sim$160\,km\,s$^{-1}$) stellar sigma. Also, the metallicity is supersolar, roughly constant within the limits of our FoV.

\subsection{NGC\,3786}
%GMOS datacube never analyzed
This galaxy receives the SABa classification, with a Seyfert\,1.9 nucleus. According to our analysis, the inner $\sim$0\farcs8 are dominated by a young and slightly more metallic SP, co-spatial with a $\sim$50\% contribution from a FC. This region is surrounded by an older, less metallic region extending up to 1\farcs0 in all directions. Also, except for border effects, the A$_{\rm V}$ and $\sigma$ are nearly constant inside our FoV. Lastly, the V map is well reproduced by a rotating disk model.

\subsection{NGC\,4939}
%GMOS datacube never analyzed (my data)
NGC\,4939 is an SAbc galaxy, harboring a Seyfert\,2 AGN. No significant traces of SPs under 2\,Gyr were detected in this galaxy. However, a bar with $\sim$4\,Gyr and $<z>_L\sim$1.5\,Z$_\odot$ can be observed, contrasting with the rest of our FoV (t$>$10\,Gyr and solar metallicity). Also, the stellar $\sigma$ is roughly constant, at 150\,km\,s$^{-1}$, and the A$_{\rm V}$ is higher in the nucleus compared to the rest of our FoV (0.6\,mag vs 0.2\,mag). Lastly, we did not detect any significant contributions of a FC except for border effects.

\subsection{NGC\,5728}
NGC\,5728 is an SABa galaxy, hosting a Seyfert\,1.9 AGN. The stellar properties inside our FoV are split in half, with the eastern region dominated by an intermediate-aged SP ($<t>_L\sim$9.3), a negligible reddening, and no contribution of a FC, whereas the western region is dominated by a much older ($<t>_L\sim$9.8) and more reddened SP (A$_{\rm V}\sim$1.0\,mag), with also a $\sim$10\% contribution of a FC. We also did not detect any velocity deviation from a disk in rotation. This MUSE datacube has already been analyzed by \citet{Shimizu+19} and \citet{Bittner+20}, who mapped the stellar populations of this object. Both authors reported the detection of a younger ring with $\sim$1\,Gyr reaching up to 5\farcs0 away from the AGN. According to \citet{Shimizu+19}, this star formation region is induced by the primary bar of this galaxy, which can be seen in our continuum image. By comparing our results with those derived by them, it is possible to see that the region with younger SPs and lower dust reddening is part of the big ring these authors detected.

\subsection{NGC\,5899}
%GMOS datacube never analyzed
NGC\,5899 is an SABc galaxy, hosting a Seyfert\,2 AGN. This galaxy is dominated by an old SP, with ages between 5 and 7\,Gyr. At the edges of our FoV, we detected some contribution of young and intermediate-age SPs, but these results may be affected by border effects. Traces of a FC were detected  in the inner 1\farcs0, contributing with up to 10\% at some spaxels. Except for statistical uncertainties, the dust reddening, stellar $\sigma$ and metallicity appear to be nearly constant within our FoV, without any discernible pattern. Lastly, we did not detect any deviation from a rotating disk in the V map.

\section{General discussion}
\label{sec:discussion}

Our sample is composed of six early-type galaxies (ETGs) and eight late-type galaxies (LTGs). We averaged the properties of early and late type galaxies separately, which are presented in Table~\ref{tab:morph}. It is possible to see that most properties of our sample are indeed correlated with their morphological type. ETGs are associated with older, less reddened SPs, with also higher velocity dispersion. LTGs, on the other hand, display a higher contribution of a FC, as well as intermediate-age and young stellar populations, coupled with higher dust reddening, as well as lower velocity dispersion.

\begin{table}
    \centering
    \begin{tabular}{lcccccccc}
    \hline
    Class & FC & xy & xi & xo & ${\rm <t>_L}$ &A$_{\rm V}$ & $\sigma$ & ${\rm <Z>_L}$\\
    \hline
    ETG &  3.8 & 3.9 & 11.4 & 80.9 & 9.7 & 0.54 & 154 & 1.32\\
    LTG &  6.0 & 5.9 & 13.1 & 75.0 & 9.6 & 0.69 & 139 & 1.33\\
    \end{tabular}
    \caption{Properties of our sample averaged by morphological type}
    \label{tab:morph}
\end{table}

\begin{figure*}
    %\centering
    \includegraphics[width=\textwidth]{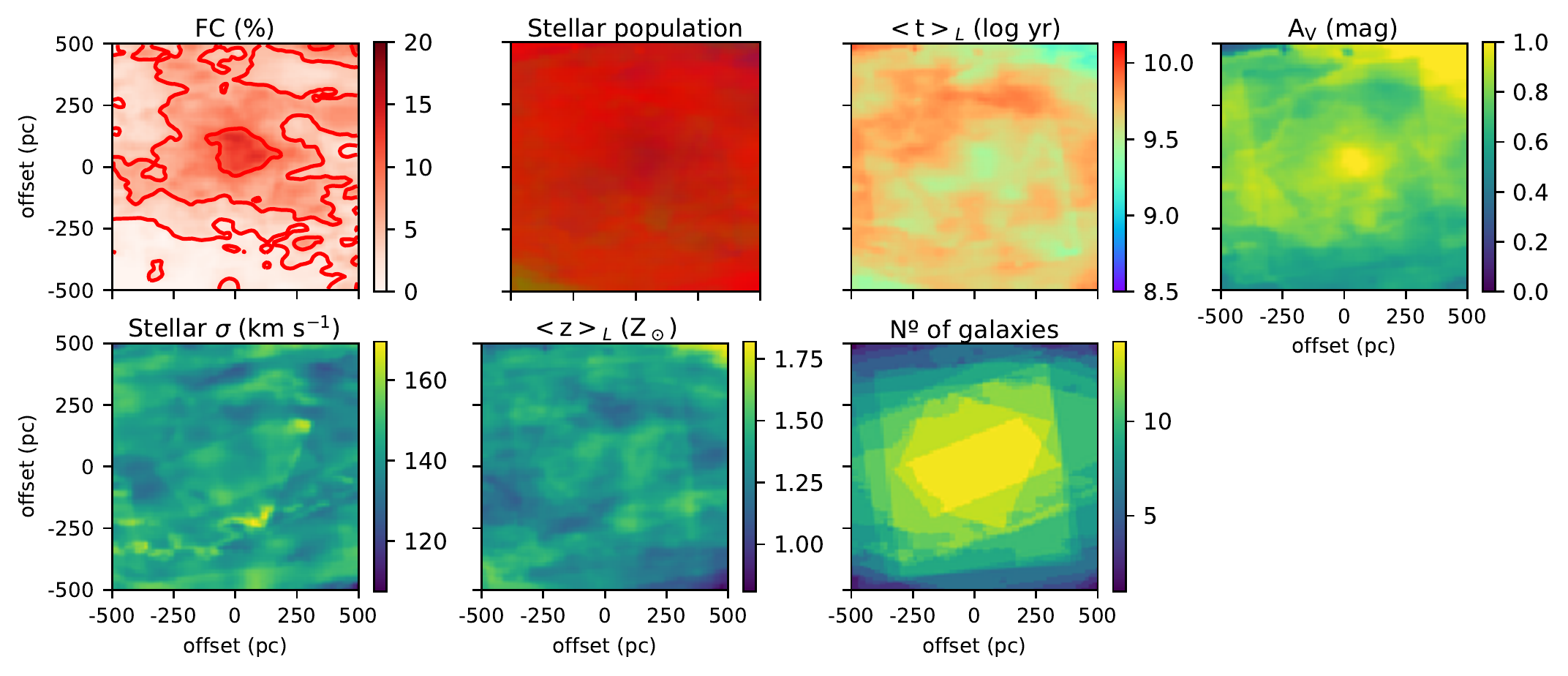}
    \caption{Averaged stellar properties of our sample. In the first line, we show, from left to right: the percent contribution from a FC, with contours drawn at 2, 5 and 10\%; an RGB image composed of young SSPs (blue), intermediate-age SSPs (green) and old SSPs (red); the logarithm of the mean stellar age, weighted by luminosity; and the dust reddening. In the second line, from left to right: the stellar $\sigma$; the average metallicity weighted by luminosity; and the number of galaxies contributing to each region.}
    \label{fig:aspv8}
\end{figure*}

\begin{figure*}
    %\centering
    \includegraphics[width=\textwidth]{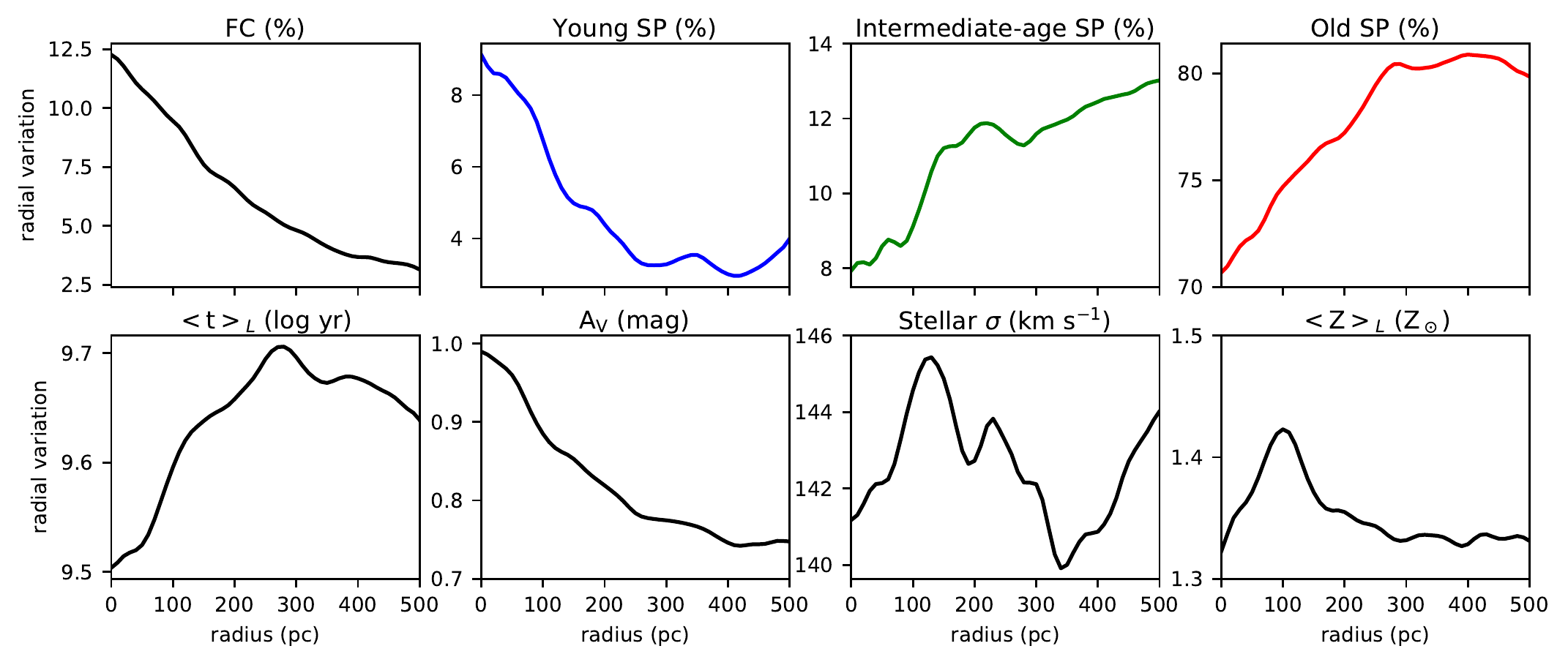}
    \caption{Radial profiles of the main properties of our sample. In the first row, from left to right we show: the percentage contribution from a featureless continuum; young SPs; intermediate-age SPs; old SPs. In the second row, we show from left to right: the logarithm of the mean stellar age, weighted by luminosity; the dust reddening; the stellar $\sigma$; the average metallicity weighted by luminosity.}
    \label{fig:aspv8rp}
\end{figure*}

In order to probe the impact of the AGN over the properties of our sample, we stacked all galaxies and mapped their average stellar properties. In order to achieve this, we converted the angular scale to the physical scale, and then fixed the luminosity peak of all galaxies to be the same. Then, we obtained position angles of the major axis, as well as axis ratios from the The Two Micron All Sky Survey \citep[2MASS][]{Skrutskie+06}, and corrected for projection effects. We fixed the major axis of the galaxies along the y direction, and the minor axis along the x axis. After that, we removed all regions with S/N<6, and averaged each individual property within each region covered by our data. Also, in order to improve the S/N at the borders of the figure, we included all MUSE galaxies up to 500\,pc. After stacking the properties, we constructed radial profiles by integrating rings with 10\,pc radii, starting from the center. These results are presented in Figs.~\ref{fig:aspv8} and \ref{fig:aspv8rp} in the shape of maps and radial profiles, respectively.

\subsection{Stellar Population}

Fig~\ref{fig:aspv8rp} shows that the contribution of the FC, as well as young SPs, is maximum closer to the AGN, decreasing outwards. The contribution of the intermediate age SP presents a different behavior, with the nucleus exhibiting a smaller fraction (8\%) if compared to the outer regions (13\%). Old SPs vary less in comparison with young and intermediate-age ones, with a minimum of 70\% in the nucleus compared with 80\% at 500\,pc. \citet{Mallmann+18} already reported a similar result, detecting an excess of young stellar populations close to the more luminous AGNs if compared to non-AGN hosts. Since our sample was obtained from data archives, we do not have a control sample to compare with. However, since we are observing the bulge of the galaxy, the existence of a star formation peak in the nucleus is very significant, as the bulge tends to be dominated by older SPs. 
\par
These results are consistent with the idea of a rise in the SFR related to the AGN. Also, combined with the result found by the simulations of \citet{Zubovas&Bourne17} and the CALIFA results of \citet{Lacerda+20}, that the AGN acts in the gas by removing or heating it, our findings support a scenario in which the starburst phase is followed by the AGN phase, but after the AGN is activated it sweeps the remaining gas in the central regions, thus quenching star formation. For this reason, AGNs would be transitioning between star-forming and quenched galaxies for two reasons: (i) part of the gas would be depleted by the excessive star formation taking place and (ii) the remaining gas would be removed/heated by the nuclear activity, thus ceasing the star formation.
\par
Also, although for more massive BHs the M-$\sigma$ relation is usually explained by coalescence \citep[as a consequence of galaxy mergers, see][]{Kormendy&Ho13,vandenBosch16}, this scenario helps explaining the M-$\sigma$ relation for less massive BHs. The SMBH would grow with the bulge by gas inflow, but after activated would quench the circumnuclear star formation, thus regulating the bulge mass. An alternative explanation is that the outflows would both enhance (positive feedback) as well as suppress (negative feedback) the star formation in different regions of the host galaxy \citep[e.g.][]{Dugan+17,Cielo+18,Mukherjee+18,AlYazeedi+21}, depending on factors such as inclination, energy and mode (quasar or radio) of the jet. We did not detect any sizeable stellar feature in the radial direction, which could be caused by outflows. All properties derived are symmetric in radius, with the exception of the dust lane in NGC\,2992, the spiral arms in NGC\,3393 and the east-west division in NGC\,5728. However, this does not rule out the hypothesis of positive nor negative feedback at larger radii, as well as in individual cases.

\subsection{Testing the robustness of our results} 
\subsubsection{Statistical Uncertainties}
\label{sec:statistical}

\citet{CF+14} estimated {\sc starlight} uncertainties by inducing $\sigma$-level errors in a CALIFA datacube of NGC\,2916. They found resulting uncertainties in age determination of 3, 9 and 9\% for the young, intermediate-age and old SP bins, respectively. \citet{Burtscher+21} also estimated {\sc starlight} errors by performing a statistical analysis over the Markov Chains used by the code, and reported much lower uncertainties, of 0.34, 0.28, 2.07 and 1.99 for their AGN sample, and 0.37, 0.62, 2.37 and 2.13\% for their control sample (in four age bins: young, young-intermediate, intermediate-old and old).
\par
In order to estimate the uncertainties in our data, we compared the stellar populations of regions smaller than the PSF. Since our data is seeing limited, we can derive an upper limit for the uncertainties of our data, by assuming all eight adjacent spaxels correspond to the same region. By performing such analysis, we were able to estimate errors of 5, 3, 5 and 6\% for the FC, xy, xi and xo, and an error of 0.1 mag for the dust reddening.
\par
Also, for the four objects in our sample, with both GMOS and MUSE data, we integrated and compared a circular region with 100\,pc in radius in both GMOS and MUSE datacubes. This comparison is presented in Table~\ref{tab:comparison}. Although the percentages vary a little, the dominant stellar population remains the same. From this analysis, we estimated errors of 3, 2, 4 and 5\% for the FC, xy, xi and xo, as well as an error of 0.2 mag for the dust reddening.

\begin{table}
    \centering
    \begin{tabular}{lcccccc}
    Object           & instrument& FC &xy & xi & xo & A$_{\rm V}$ \\\hline
    \multirow{2}{*}{NGC\,1068}&GMOS&24.1&3.0 &42.4&30.4&0.17\\
                              &MUSE&18.6&1.2 &44.8&36.4&0.26\\\hline
    \multirow{2}{*}{NGC\,2992}&GMOS&5.9 &1.0 &3.2 &89.8&2.32\\
                              &MUSE&5.6 &0.0 &0.1 &94.3&2.15\\\hline   
    \multirow{2}{*}{NGC\,3393}&GMOS&0.6 &1.4 &1.3 &96.7&0.79\\
                              &MUSE&0.0 &0.20&0.0 &99.8&0.58\\\hline
    \multirow{2}{*}{NGC\,5728}&GMOS&1.6 &1.5 &22.2&74.6&0.18\\
                              &MUSE&0.8 &0.0 &27.9&71.3&0.38\\\hline
    \end{tabular}
    \caption{Comparison of integrated results for objects having both GMOS and MUSE datacubes.}
    \label{tab:comparison}
\end{table}

\subsubsection{Degeneracies}
There are three sources of degeneracy that could be playing a role in our results. The first is that, according to \citet{CF+04}, a reddened young starburst (t$\leq$5\,Myr) is indistinguishable from a FC. \citet{Burtscher+21} tested the capability of {\sc starlight} to distinguish between a FC and young SSPs, and the code proved capable of separating blue light originating from the FC from blue light from young SPs. Also, by comparing the SFR derived from {\sc starlight} with the SFR derived from the emission lines in their sample, \citet{riffel+21} reported that they closely reproduce each other.
\par
In order to verify if part of the young SPs in our sample can be attributed to a FC emitted by the AGN, rather than a younger population, we performed the stellar synthesis without SSPs younger than 30\,Myr. If very young SSPs were being used by {\sc starlight} to reproduce the FC component from the AGN, after removing these SSPs from the library would force {\sc starlight} to use the FC component instead. This test is not perfect, since it would also cause {\sc starlight} to reproduce any real very young SPs with a FC. Except for small variations due to the S/N, our results remained the same. This not only supports the presence of a young SP in the nucleus, it also puts a constraint on the age of this starburst, as it cannot be reproduced by a power law, and thus this starburst is older than 5\,Myr, since SPs below this threshold have no stellar absorptions besides those from Hydrogen, and thus they would be replaced by a FC.
\par
Past studies have suggested that AGNs operate in cycles, with each flickering event lasting typically $\sim10^5$\,yr \citep{Schawinski+15,King&Nixon15}. This timespan is much shorter than the lower threshold imposed by our analysis, which implies that the nuclear SPs we detected are older than the ongoing nuclear activity cycle. This, in turn, means that these young SPs cannot be a consequence of the ongoing AGN activity, but either the nuclear activity is a consequence of the starburst, or these two events have a common origin, with the star formation taking place first.
\par
The second is the so called age-metallicity degeneracy \citep{Worthey94}. According to this degeneracy, an old metal-poor stellar population is almost identical to a younger, and metal-rich one. This degeneracy is harder to break, since the indices sensitive to ages, such as H$\alpha$ and H$\beta$ are dominated by emission lines from the AGN. However, as can be seen in Fig~\ref{fig:aspv8rp}, the metallicity is almost constant within our FoV, with a value close to 1.4\,Z$_\odot$. This suggests that the age-metallicity degeneracy does not play a role in our case.
\par
The last spectral degeneracy present in stellar population syntheses is the degeneracy between dust extinction and stellar age. However, since extinction only affects the continuum, and not stellar absorptions, the effect is much more dramatic when dealing with SEDs, compared to our fitting procedure, which takes into account these absorptions. Even though this effect is small, it is non-negligible, and can be seen as vertical bands showing an anti-correlation between dust reddening and metallicity in at least three objects (NGC\,2110, NGC\,3516, NGC\,5899). This happens because the instrumental fingerprints do not have absorption features, and thus they alter the continuum shape of the spectra, which {\sc starlight} tries to compensate by mixing stellar populations and dust reddening in different proportions. Based on these bands, we found errors compatible with the ones derived in Section~\ref{sec:statistical}. The only exception is NGC\,2110, which changes the stellar populations by a factor of up to 80\% in some spaxels. For this reason is important to construct radial profiles, since such features are diluted along different radii, and thus do not alter these results.

\subsubsection{Comparison with previous works}
Also, in order to test the consistency of our results, we compared the four objects in common with the sample of \citet{Burtscher+21}. They analyzed the integrated light from a 1\farcs8$\times$1\farcs8 region of NGC\,2110, NGC\,2992, NGC\,3081, NGC\,5728, among five other AGNs, six star-forming and 12 passive galaxies. The authors also employed {\sc starlight} code, but used \citet{BC03} library of models instead. Also, they binned their stellar populations into four categories, xy (t$<$0.0316\,Gyr), xyi (0.0316$<$t$<$0.316\,Gyr), xio (0.316$<$t$<$3.16\,Gyr) and xo (t$>$3.16\,Gyr), and did not include a FC, with the argument that their AGN sample is composed mainly of Seyferts\,2, and also the lack of correlation between the flux in the blue continuum (at 4000-4200$\r{A}$) with the hard X-ray flux from the BAT-105 month survey \citep{Oh+18}. In order to properly compare our results to theirs, we extracted the inner 1\farcs8$\times$1\farcs8 and derived an average stellar population weighted by the light continuum. This comparison is presented in Table~\ref{tab:leonard}. Our results agree that this 1\farcs8$\times$1\farcs8 region of the four galaxies in common are dominated in light by old SPs. Also, the reddening is consistent, pointing towards NGC\,2992 having the highest dust reddening, followed by NGC\,2110, with NGC\,3081 and NGC\,5728 having the lowest values in the sample. However, a few differences need to be noted: we found for NGC\,2110 significantly contributions from both young and intermediate-age SPs, while also detecting a FC related to the AGN, whereas they found only a 7\% contribution coming from young SPs. Also, they found a 5.7\% contribution from young SPs in NGC\,3081 and 4.0\% in NGC\,5728, which we did not detect. In the case of NGC\,5728, however, this could be linked to a 1.9\% FC contribution that we detected. Also, these differences could be caused by the fact that their wavelength range was larger than ours, which can play a decisive role especially at bluer wavelengths. This comparison highlights that our results are consistent compared to different studies using similar methods.

\begin{table*}
\centering
    \begin{tabular}{c@{\hskip 0.5in}ccccc@{\hskip 0.5in}ccccc}
              & \multicolumn{4}{c}{Burtscher et al.} && \multicolumn{5}{c}{This work}\\
    Object    & A$_{\rm V}$ & xy & xyi & xio & xo    & A$_{\rm V}$ & FC & xy & xi & xo\\
    NGC\,2110 &  1.18 & 7.0 & 0.0 & 0.0 & 93.0  & 0.8 & 14.8& 16.8 & 20.8& 47.6\\
    NGC\,2992 &  1.87 & 0.0 & 0.0 &26.6 & 73.4  & 1.4 & 1.5 & 0.1  & 0.1 & 98.3\\
    NGC\,3081 &  0.51 & 5.7 & 0.0 & 2.5 & 91.8  & 0.6 & 0.0 & 0.0  & 0.1 & 99.9\\
    NGC\,5728 &  0.83 & 4.0 & 0.0 &20.2 & 75.8  & 0.5 & 1.9 & 0.1  & 33.6 & 64.4\\
    \end{tabular}
\caption{Comparison between the results obtained by \citet{Burtscher+21} and this work for the inner 1\farcs8$\times$1\farcs8.}
\label{tab:leonard}
\end{table*}

\subsection{Stellar Kinematics}
The stellar velocity fields of our galaxies are well reproduced by assuming that part of the stars are in circular motion around the nucleus, whereas the remaining stars follow random motions, which can be observed from the velocity dispersion. Besides, we did not detect further deviations from circular motions.
\par
The stellar kinematics for part of our sample were previously analyzed by \citet{RiffelRA+17}, based on NIR K-band datacubes. They also found that the observed velocities are well reproduced by rotating disc models. Also, for two galaxies in our sample (Mrk\,1066 and NGC\,5899), they found an S-shape zero velocity line. For Mrk\,1066, we could confirm the existence of this structure. For NGC\,5899, on the other hand, we did not detect such feature, even though our FoV if much bigger than theirs. This difference might be caused by statistical uncertainties in any of the two datasets, since this feature is more evident in the edges of their datacube. Also, their sample did not include NGC\,2992, for which we also detected an S-shaped zero velocity line.
\par
Another important result that we found, is that the velocity dispersion is mainly flat inside our FoV, varying by less than 6\,km\,s$^{-1}$. From all the galaxies in our sample, NGC\,1052, NGC\,3081, NGC\,3393 showed signs of $\sigma$ peaks, and only Mrk\,607 and NGC\,1068 showed prominent $\sigma$ drops in the central regions. Since we detected a difference in the stellar populations in the central region, a corresponding change in velocity dispersion would help shedding some light into the kinematic properties of the gas that triggered the starburst. 
\par
A similar result has already been found by \cite{Lin+18}, who analyzed eight AGNs and five control galaxies. Having detected a nuclear stellar light excess in their AGNs compared to their control sample, which they attributed to a younger stellar population, they also did not detect a significant change in their stellar $\sigma$ in this region. They argued that the lack of any significant change in $\sigma$ associated with this luminosity excess either implies the existence of pseudo-bulges, in which the intrinsic dispersion increases towards the centre, or that the fraction of young stellar populations is too small to dominate over the bulge.

\subsection{Dust reddening}
We also found that, on average, the dust reddening is slightly higher in the inner 200\,pc, with a 1.0\,mag peak in the nucleus, compared to an average of 0.75\,mag in the external regions. This is also consistent with a density increase in the inner regions, both in gas and dust, which could also be associated to the infall of gas, also triggering the observed star-formation peak. 
\par
However, this difference is very small, especially if compared to the dust torus of the AGN, which is higher than 20 magnitudes for all Seyfert\,2 galaxies in \citet{Burtscher+16} sample. This highlights that the extinction to the stellar population is related to larger scale properties, and unconnected to the gas column and extinction closer to the AGN.

\section{Conclusions}
\label{sec:conclusions}
 We have used  GMOS and MUSE integral field unit data for a sample of 14 Seyfert galaxies in order to derive spatially resolved age, metallicity and kinematic properties of the sources. Our main results can be summarized as follows:

% We have used  GMOS and MUSE integral field unit archival data for a sample of 14 Seyfert galaxies. We performed stellar population synthesis technique, and derived age, metallicity and kinematic properties. After deriving these maps individually, we stacked all objects in our sample. Our main results can be summarized as follows:
\begin{itemize}
    \item The stellar properties of our sample are correlated with their morphological types. Whereas elliptical/lenticular galaxies are associated with old, less reddened SPs, with also high velocity dispersion, spirals exhibit higher fraction of FC contribution, as well as young and intermediate-age SSPs, combined with lower velocity dispersion and higher dust reddening;
    \item After averaging the stellar properties of our galaxies, we were able to detect a peak in the contribution of young SPs in the central 100\,pc closer to the AGN, if compared to the rest of our FoV;
    \item  The behaviour of the intermediate-age SPs is opposite to the young SPs, with a lower value in the nucleus. Also, the fraction of old SPs vary very little within our FoV;
    \item We were able to detect a rise in the dust reddening closer to the nucleus, peaking at 1.0\,mag. This dust, however, is associated with galaxy-wide properties rather than the dust torus;
    \item There is no significant change in the stellar velocity dispersion associated with the changes in the SP. This suggests either the existence of pseudo-bulges, in which the intrinsic dispersion increases towards the centre, or that the fraction of young stellar populations is too small to dominate over the bulge;
\end{itemize}

Our results, combined with previous works \citep{Mallmann+18,Burtscher+21}, indicate that AGNs are associated with recent star formation. This suggests that the AGNs have been triggered by a recent supply of gas that has previously triggered a star formation episode in the central region of their AGN host galaxies.

\section*{Acknowledgements}

LGDH thanks CNPq, LNA and SHAO for funding and support. RR thanks CNPq (311223/2020-6), CAPES and FAPERGS
(16/2551-0000251-7 and 19/2551-0001750-2). ARA thanks Conselho Nacional de Desenvolvimento Cient\'ifico e Tecnol\'ogico (CNPq) for partial support to this work through grant 312036/2019-1. RAR acknowledges partial financial support from CNPq and FAPERGS. This research has made use of the services of the ESO Science Archive Facility. Based on observations collected at the European Southern Observatory under ESO programmes 094.B-0298(A), 094.B-0321(A), 097.B-0640(A), 098.B-0551(A) and 099.B-0242(B).

\section*{Data Availability}

The GMOS data underlying this article are available in the Gemini Observatory Archive\footnote{https://archive.gemini.edu/searchform}, and can be accessed with projects number GN-2013A-Q-61, GS-2013B-Q-20,  GN-2014B-Q-87, GN-2017B-Q-44 and GS-2018A-Q-225. The MUSE data underlying this article are available in the ESO Science Archive Facility\footnote{http://archive.eso.org/cms.html}, and can be accessed with projects number 094.B-0298(A), 094.B-0321(A), 097.B-0640(A), 098.B-0551(A) and 099.B-0242(B). 

\bibliographystyle{mnras}
\bibliography{main}

\begin{thebibliography}{}
\makeatletter
\relax
\def\mn@urlcharsother{\let\do\@makeother \do\$\do\&\do\#\do\^\do\_\do\%\do\~}
\def\mn@doi{\begingroup\mn@urlcharsother \@ifnextchar [ {\mn@doi@}
  {\mn@doi@[]}}
\def\mn@doi@[#1]#2{\def\@tempa{#1}\ifx\@tempa\@empty \href
  {http://dx.doi.org/#2} {doi:#2}\else \href {http://dx.doi.org/#2} {#1}\fi
  \endgroup}
\def\mn@eprint#1#2{\mn@eprint@#1:#2::\@nil}
\def\mn@eprint@arXiv#1{\href {http://arxiv.org/abs/#1} {{\tt arXiv:#1}}}
\def\mn@eprint@dblp#1{\href {http://dblp.uni-trier.de/rec/bibtex/#1.xml}
  {dblp:#1}}
\def\mn@eprint@#1:#2:#3:#4\@nil{\def\@tempa {#1}\def\@tempb {#2}\def\@tempc
  {#3}\ifx \@tempc \@empty \let \@tempc \@tempb \let \@tempb \@tempa \fi \ifx
  \@tempb \@empty \def\@tempb {arXiv}\fi \@ifundefined
  {mn@eprint@\@tempb}{\@tempb:\@tempc}{\expandafter \expandafter \csname
  mn@eprint@\@tempb\endcsname \expandafter{\@tempc}}}

\bibitem[\protect\citeauthoryear{{Ajello}, {Alexander}, {Greiner}, {Madejski},
  {Gehrels}  \& {Burlon}}{{Ajello} et~al.}{2012}]{Ajello+12}
{Ajello} M.,  {Alexander} D.~M.,  {Greiner} J.,  {Madejski} G.~M.,  {Gehrels}
  N.,   {Burlon} D.,  2012, \mn@doi [\apj] {10.1088/0004-637X/749/1/21}, \href
  {https://ui.adsabs.harvard.edu/abs/2012ApJ...749...21A} {749, 21}

\bibitem[\protect\citeauthoryear{{Al Yazeedi}, {Katkov}, {Gelfand},
  {Wylezalek}, {Zakamska}  \& {Liu}}{{Al Yazeedi} et~al.}{2021}]{AlYazeedi+21}
{Al Yazeedi} A.,  {Katkov} I.~Y.,  {Gelfand} J.~D.,  {Wylezalek} D.,
  {Zakamska} N.~L.,   {Liu} W.,  2021, \mn@doi [\apj]
  {10.3847/1538-4357/abf5e1}, \href
  {https://ui.adsabs.harvard.edu/abs/2021ApJ...916..102A} {916, 102}

\bibitem[\protect\citeauthoryear{{Audibert}, {Riffel}, {Sales}, {Pastoriza}  \&
  {Ruschel-Dutra}}{{Audibert} et~al.}{2017}]{Audibert+17}
{Audibert} A.,  {Riffel} R.,  {Sales} D.~A.,  {Pastoriza} M.~G.,
  {Ruschel-Dutra} D.,  2017, \mn@doi [\mnras] {10.1093/mnras/stw2477}, \href
  {https://ui.adsabs.harvard.edu/abs/2017MNRAS.464.2139A} {464, 2139}

\bibitem[\protect\citeauthoryear{{Bertola}, {Bettoni}, {Danziger}, {Sadler},
  {Sparke}  \& {de Zeeuw}}{{Bertola} et~al.}{1991}]{Bertola+91}
{Bertola} F.,  {Bettoni} D.,  {Danziger} J.,  {Sadler} E.,  {Sparke} L.,   {de
  Zeeuw} T.,  1991, \mn@doi [\apj] {10.1086/170058}, \href
  {http://adsabs.harvard.edu/abs/1991ApJ...373..369B} {373, 369}

\bibitem[\protect\citeauthoryear{{Bischetti} et~al.,}{{Bischetti}
  et~al.}{2017}]{Bischetti+17}
{Bischetti} M.,  et~al., 2017, \mn@doi [\aap] {10.1051/0004-6361/201629301},
  \href {https://ui.adsabs.harvard.edu/abs/2017A&A...598A.122B} {598, A122}

\bibitem[\protect\citeauthoryear{{Bittner} et~al.,}{{Bittner}
  et~al.}{2020}]{Bittner+20}
{Bittner} A.,  et~al., 2020, \mn@doi [\aap] {10.1051/0004-6361/202038450},
  \href {https://ui.adsabs.harvard.edu/abs/2020A&A...643A..65B} {643, A65}

\bibitem[\protect\citeauthoryear{{Bruzual} \& {Charlot}}{{Bruzual} \&
  {Charlot}}{2003}]{BC03}
{Bruzual} G.,  {Charlot} S.,  2003, \mn@doi [\mnras]
  {10.1046/j.1365-8711.2003.06897.x}, \href
  {http://adsabs.harvard.edu/abs/2003MNRAS.344.1000B} {344, 1000}

\bibitem[\protect\citeauthoryear{{Burtscher} et~al.,}{{Burtscher}
  et~al.}{2016}]{Burtscher+16}
{Burtscher} L.,  et~al., 2016, \mn@doi [\aap] {10.1051/0004-6361/201527575},
  \href {https://ui.adsabs.harvard.edu/abs/2016A&A...586A..28B} {586, A28}

\bibitem[\protect\citeauthoryear{{Burtscher} et~al.,}{{Burtscher}
  et~al.}{2021}]{Burtscher+21}
{Burtscher} L.,  et~al., 2021, arXiv e-prints, \href
  {https://ui.adsabs.harvard.edu/abs/2021arXiv210505309B} {p. arXiv:2105.05309}

\bibitem[\protect\citeauthoryear{{Cai}, {Zhao}, {Zhang}, {Bai}  \& {Liu}}{{Cai}
  et~al.}{2020}]{Cai+20}
{Cai} W.,  {Zhao} Y.,  {Zhang} H.-X.,  {Bai} J.-M.,   {Liu} H.-T.,  2020,
  \mn@doi [\apj] {10.3847/1538-4357/abb81c}, \href
  {https://ui.adsabs.harvard.edu/abs/2020ApJ...903...58C} {903, 58}

\bibitem[\protect\citeauthoryear{{Calzetti}, {Armus}, {Bohlin}, {Kinney},
  {Koornneef}  \& {Storchi-Bergmann}}{{Calzetti} et~al.}{2000}]{Calzetti+00}
{Calzetti} D.,  {Armus} L.,  {Bohlin} R.~C.,  {Kinney} A.~L.,  {Koornneef} J.,
   {Storchi-Bergmann} T.,  2000, \mn@doi [\apj] {10.1086/308692}, \href
  {http://adsabs.harvard.edu/abs/2000ApJ...533..682C} {533, 682}

\bibitem[\protect\citeauthoryear{{Chen}, {Liang}, {Hammer}, {Prugniel},
  {Zhong}, {Rodrigues}, {Zhao}  \& {Flores}}{{Chen} et~al.}{2010}]{Chen+10}
{Chen} X.~Y.,  {Liang} Y.~C.,  {Hammer} F.,  {Prugniel} P.,  {Zhong} G.~H.,
  {Rodrigues} M.,  {Zhao} Y.~H.,   {Flores} H.,  2010, \mn@doi [\aap]
  {10.1051/0004-6361/200913894}, \href
  {http://adsabs.harvard.edu/abs/2010A%26A...515A.101C} {515, A101}

\bibitem[\protect\citeauthoryear{{Cid Fernandes}, {Gu}, {Melnick}, {Terlevich},
  {Terlevich}, {Kunth}, {Rodrigues Lacerda}  \& {Joguet}}{{Cid Fernandes}
  et~al.}{2004}]{CF+04}
{Cid Fernandes} R.,  {Gu} Q.,  {Melnick} J.,  {Terlevich} E.,  {Terlevich} R.,
  {Kunth} D.,  {Rodrigues Lacerda} R.,   {Joguet} B.,  2004, \mn@doi [\mnras]
  {10.1111/j.1365-2966.2004.08321.x}, \href
  {http://adsabs.harvard.edu/abs/2004MNRAS.355..273C} {355, 273}

\bibitem[\protect\citeauthoryear{{Cid Fernandes}, {Mateus}, {Sodr{\'e}},
  {Stasi{\'n}ska}  \& {Gomes}}{{Cid Fernandes} et~al.}{2005}]{CF+05}
{Cid Fernandes} R.,  {Mateus} A.,  {Sodr{\'e}} L.,  {Stasi{\'n}ska} G.,
  {Gomes} J.~M.,  2005, \mn@doi [\mnras] {10.1111/j.1365-2966.2005.08752.x},
  \href {http://adsabs.harvard.edu/abs/2005MNRAS.358..363C} {358, 363}

\bibitem[\protect\citeauthoryear{{Cid Fernandes} et~al.,}{{Cid Fernandes}
  et~al.}{2013}]{CF+13}
{Cid Fernandes} R.,  et~al., 2013, \mn@doi [\aap]
  {10.1051/0004-6361/201220616}, \href
  {http://adsabs.harvard.edu/abs/2013A%26A...557A..86C} {557, A86}

\bibitem[\protect\citeauthoryear{{Cid Fernandes} et~al.,}{{Cid Fernandes}
  et~al.}{2014}]{CF+14}
{Cid Fernandes} R.,  et~al., 2014, \mn@doi [\aap]
  {10.1051/0004-6361/201321692}, \href
  {https://ui.adsabs.harvard.edu/abs/2014A&A...561A.130C} {561, A130}

\bibitem[\protect\citeauthoryear{{Cielo}, {Bieri}, {Volonteri}, {Wagner}  \&
  {Dubois}}{{Cielo} et~al.}{2018}]{Cielo+18}
{Cielo} S.,  {Bieri} R.,  {Volonteri} M.,  {Wagner} A.~Y.,   {Dubois} Y.,
  2018, \mn@doi [\mnras] {10.1093/mnras/sty708}, \href
  {https://ui.adsabs.harvard.edu/abs/2018MNRAS.477.1336C} {477, 1336}

\bibitem[\protect\citeauthoryear{{Colina}, {Fricke}, {Kollatschny}  \&
  {Perryman}}{{Colina} et~al.}{1987}]{Colina+87}
{Colina} L.,  {Fricke} K.~J.,  {Kollatschny} W.,   {Perryman} M.~A.~C.,  1987,
  \aap, \href {https://ui.adsabs.harvard.edu/abs/1987A&A...178...51C} {178, 51}

\bibitem[\protect\citeauthoryear{{Crenshaw} \& {Kraemer}}{{Crenshaw} \&
  {Kraemer}}{2000}]{Crenshaw&Kraemer00}
{Crenshaw} D.~M.,  {Kraemer} S.~B.,  2000, \mn@doi [\apj] {10.1086/308570},
  \href {https://ui.adsabs.harvard.edu/abs/2000ApJ...532..247C} {532, 247}

\bibitem[\protect\citeauthoryear{{Dahmer-Hahn}, {Riffel},
  {Rodr{\'{\i}}guez-Ardila}, {Martins}, {Kehrig}, {Heckman}, {Pastoriza}  \&
  {Dametto}}{{Dahmer-Hahn} et~al.}{2018}]{luisgdh+18}
{Dahmer-Hahn} L.~G.,  {Riffel} R.,  {Rodr{\'{\i}}guez-Ardila} A.,  {Martins}
  L.~P.,  {Kehrig} C.,  {Heckman} T.~M.,  {Pastoriza} M.~G.,   {Dametto} N.~Z.,
   2018, \mn@doi [\mnras] {10.1093/mnras/sty515}, \href
  {http://adsabs.harvard.edu/abs/2018MNRAS.476.4459D} {476, 4459}

\bibitem[\protect\citeauthoryear{{Dahmer-Hahn} et~al.,}{{Dahmer-Hahn}
  et~al.}{2019}]{luisgdh+19a}
{Dahmer-Hahn} L.~G.,  et~al., 2019, \mn@doi [\mnras] {10.1093/mnras/sty3051},
  \href {https://ui.adsabs.harvard.edu/\#abs/2019MNRAS.482.5211D} {482, 5211}

\bibitem[\protect\citeauthoryear{{Diniz}, {Riffel}, {Storchi-Bergmann}  \&
  {Riffel}}{{Diniz} et~al.}{2019}]{Diniz+19}
{Diniz} M.~R.,  {Riffel} R.~A.,  {Storchi-Bergmann} T.,   {Riffel} R.,  2019,
  \mn@doi [\mnras] {10.1093/mnras/stz1329}, \href
  {https://ui.adsabs.harvard.edu/abs/2019MNRAS.487.3958D} {487, 3958}

\bibitem[\protect\citeauthoryear{{Dugan}, {Gaibler}  \& {Silk}}{{Dugan}
  et~al.}{2017}]{Dugan+17}
{Dugan} Z.,  {Gaibler} V.,   {Silk} J.,  2017, \mn@doi [\apj]
  {10.3847/1538-4357/aa7566}, \href
  {https://ui.adsabs.harvard.edu/abs/2017ApJ...844...37D} {844, 37}

\bibitem[\protect\citeauthoryear{{Ellison}, {Teimoorinia}, {Rosario}  \&
  {Mendel}}{{Ellison} et~al.}{2016}]{Ellison+16}
{Ellison} S.~L.,  {Teimoorinia} H.,  {Rosario} D.~J.,   {Mendel} J.~T.,  2016,
  \mn@doi [\mnras] {10.1093/mnrasl/slw012}, \href
  {http://adsabs.harvard.edu/abs/2016MNRAS.458L..34E} {458, L34}

\bibitem[\protect\citeauthoryear{{Erroz-Ferrer} et~al.,}{{Erroz-Ferrer}
  et~al.}{2019}]{ErrozFerrer+19}
{Erroz-Ferrer} S.,  et~al., 2019, \mn@doi [\mnras] {10.1093/mnras/stz194},
  \href {https://ui.adsabs.harvard.edu/abs/2019MNRAS.484.5009E} {484, 5009}

\bibitem[\protect\citeauthoryear{{Ferland} \& {Netzer}}{{Ferland} \&
  {Netzer}}{1983}]{Ferland&Netzer83}
{Ferland} G.~J.,  {Netzer} H.,  1983, \mn@doi [\apj] {10.1086/160577}, \href
  {http://adsabs.harvard.edu/abs/1983ApJ...264..105F} {264, 105}

\bibitem[\protect\citeauthoryear{{Ferrarese} \& {Merritt}}{{Ferrarese} \&
  {Merritt}}{2000}]{Ferrarese&Merritt00}
{Ferrarese} L.,  {Merritt} D.,  2000, \mn@doi [\apjl] {10.1086/312838}, \href
  {http://adsabs.harvard.edu/abs/2000ApJ...539L...9F} {539, L9}

\bibitem[\protect\citeauthoryear{{Ferr{\'e}-Mateu}, {Vazdekis}, {Trujillo},
  {S{\'a}nchez-Bl{\'a}zquez}, {Ricciardelli}  \& {de la
  Rosa}}{{Ferr{\'e}-Mateu} et~al.}{2012}]{Ferre-Mateu+12}
{Ferr{\'e}-Mateu} A.,  {Vazdekis} A.,  {Trujillo} I.,
  {S{\'a}nchez-Bl{\'a}zquez} P.,  {Ricciardelli} E.,   {de la Rosa} I.~G.,
  2012, \mn@doi [\mnras] {10.1111/j.1365-2966.2012.20897.x}, \href
  {https://ui.adsabs.harvard.edu/abs/2012MNRAS.423..632F} {423, 632}

\bibitem[\protect\citeauthoryear{{Forbes}, {Georgakakis}  \& {Brodie}}{{Forbes}
  et~al.}{2001}]{Forbes+01}
{Forbes} D.~A.,  {Georgakakis} A.~E.,   {Brodie} J.~P.,  2001, \mn@doi [\mnras]
  {10.1046/j.1365-8711.2001.04543.x}, \href
  {http://adsabs.harvard.edu/abs/2001MNRAS.325.1431F} {325, 1431}

\bibitem[\protect\citeauthoryear{{Freitas} et~al.,}{{Freitas}
  et~al.}{2018}]{Freitas+18}
{Freitas} I.~C.,  et~al., 2018, \mn@doi [\mnras] {10.1093/mnras/sty303}, \href
  {https://ui.adsabs.harvard.edu/abs/2018MNRAS.476.2760F} {476, 2760}

\bibitem[\protect\citeauthoryear{{Gebhardt} et~al.,}{{Gebhardt}
  et~al.}{2000}]{Gebhardt+00}
{Gebhardt} K.,  et~al., 2000, \mn@doi [\apjl] {10.1086/312840}, \href
  {http://adsabs.harvard.edu/abs/2000ApJ...539L..13G} {539, L13}

\bibitem[\protect\citeauthoryear{{Goddard} et~al.,}{{Goddard}
  et~al.}{2017}]{Goddard+17}
{Goddard} D.,  et~al., 2017, \mn@doi [\mnras] {10.1093/mnras/stw3371}, \href
  {https://ui.adsabs.harvard.edu/abs/2017MNRAS.466.4731G} {466, 4731}

\bibitem[\protect\citeauthoryear{{Gonz{\'a}lez Delgado}, {Cervi{\~n}o},
  {Martins}, {Leitherer}  \& {Hauschildt}}{{Gonz{\'a}lez Delgado}
  et~al.}{2005}]{GonzalezDelgado+05}
{Gonz{\'a}lez Delgado} R.~M.,  {Cervi{\~n}o} M.,  {Martins} L.~P.,  {Leitherer}
  C.,   {Hauschildt} P.~H.,  2005, \mn@doi [\mnras]
  {10.1111/j.1365-2966.2005.08692.x}, \href
  {https://ui.adsabs.harvard.edu/abs/2005MNRAS.357..945G} {357, 945}

\bibitem[\protect\citeauthoryear{{Heckman}}{{Heckman}}{1980}]{Heckman80}
{Heckman} T.~M.,  1980, \aap, \href
  {http://adsabs.harvard.edu/abs/1980A%26A....87..152H} {87, 152}

\bibitem[\protect\citeauthoryear{{Ho}, {Filippenko}, {Sargent}  \& {Peng}}{{Ho}
  et~al.}{1997}]{Ho+97}
{Ho} L.~C.,  {Filippenko} A.~V.,  {Sargent} W.~L.~W.,   {Peng} C.~Y.,  1997,
  \mn@doi [\apjs] {10.1086/313042}, \href
  {http://adsabs.harvard.edu/abs/1997ApJS..112..391H} {112, 391}

\bibitem[\protect\citeauthoryear{{Hopkins} \& {Quataert}}{{Hopkins} \&
  {Quataert}}{2010}]{Hopkins&Quataert10}
{Hopkins} P.~F.,  {Quataert} E.,  2010, \mn@doi [\mnras]
  {10.1111/j.1365-2966.2010.17064.x}, \href
  {https://ui.adsabs.harvard.edu/abs/2010MNRAS.407.1529H} {407, 1529}

\bibitem[\protect\citeauthoryear{{Izumi}, {Wada}, {Fukushige}, {Hamamura}  \&
  {Kohno}}{{Izumi} et~al.}{2018}]{Izumi+18}
{Izumi} T.,  {Wada} K.,  {Fukushige} R.,  {Hamamura} S.,   {Kohno} K.,  2018,
  \mn@doi [The Astrophysical Journal] {10.3847/1538-4357/aae20b}, \href
  {https://ui.adsabs.harvard.edu/abs/2018ApJ...867...48I} {867, 48}

\bibitem[\protect\citeauthoryear{{Kauffmann} et~al.,}{{Kauffmann}
  et~al.}{2003}]{Kauffmann+03}
{Kauffmann} G.,  et~al., 2003, \mn@doi [\mnras]
  {10.1111/j.1365-2966.2003.07154.x}, \href
  {http://adsabs.harvard.edu/abs/2003MNRAS.346.1055K} {346, 1055}

\bibitem[\protect\citeauthoryear{{King} \& {Nixon}}{{King} \&
  {Nixon}}{2015}]{King&Nixon15}
{King} A.,  {Nixon} C.,  2015, \mn@doi [\mnras] {10.1093/mnrasl/slv098}, \href
  {https://ui.adsabs.harvard.edu/abs/2015MNRAS.453L..46K} {453, L46}

\bibitem[\protect\citeauthoryear{{Koratkar} \& {Blaes}}{{Koratkar} \&
  {Blaes}}{1999}]{Koratkar&Blaes99}
{Koratkar} A.,  {Blaes} O.,  1999, \mn@doi [\pasp] {10.1086/316294}, \href
  {http://adsabs.harvard.edu/abs/1999PASP..111....1K} {111, 1}

\bibitem[\protect\citeauthoryear{{Kormendy} \& {Ho}}{{Kormendy} \&
  {Ho}}{2013}]{Kormendy&Ho13}
{Kormendy} J.,  {Ho} L.~C.,  2013, \mn@doi [\araa]
  {10.1146/annurev-astro-082708-101811}, \href
  {http://adsabs.harvard.edu/abs/2013ARA%26A..51..511K} {51, 511}

\bibitem[\protect\citeauthoryear{{Lacerda}, {S{\'a}nchez}, {Cid Fernandes},
  {L{\'o}pez-Cob{\'a}}, {Espinosa-Ponce}  \& {Galbany}}{{Lacerda}
  et~al.}{2020}]{Lacerda+20}
{Lacerda} E. A.~D.,  {S{\'a}nchez} S.~F.,  {Cid Fernandes} R.,
  {L{\'o}pez-Cob{\'a}} C.,  {Espinosa-Ponce} C.,   {Galbany} L.,  2020, \mn@doi
  [\mnras] {10.1093/mnras/staa008}, \href
  {https://ui.adsabs.harvard.edu/abs/2020MNRAS.492.3073L} {492, 3073}

\bibitem[\protect\citeauthoryear{{Leslie}, {Kewley}, {Sanders}  \&
  {Lee}}{{Leslie} et~al.}{2016}]{Leslie+16}
{Leslie} S.~K.,  {Kewley} L.~J.,  {Sanders} D.~B.,   {Lee} N.,  2016, \mn@doi
  [\mnras] {10.1093/mnrasl/slv135}, \href
  {http://adsabs.harvard.edu/abs/2016MNRAS.455L..82L} {455, L82}

\bibitem[\protect\citeauthoryear{{Lin} et~al.,}{{Lin} et~al.}{2018}]{Lin+18}
{Lin} M.-Y.,  et~al., 2018, \mn@doi [\mnras] {10.1093/mnras/stx2618}, \href
  {https://ui.adsabs.harvard.edu/abs/2018MNRAS.473.4582L} {473, 4582}

\bibitem[\protect\citeauthoryear{{Lutz} et~al.,}{{Lutz} et~al.}{2010}]{Lutz+10}
{Lutz} D.,  et~al., 2010, \mn@doi [\apj] {10.1088/0004-637X/712/2/1287}, \href
  {http://adsabs.harvard.edu/abs/2010ApJ...712.1287L} {712, 1287}

\bibitem[\protect\citeauthoryear{{Madau} \& {Dickinson}}{{Madau} \&
  {Dickinson}}{2014}]{Madau&Dickinson14}
{Madau} P.,  {Dickinson} M.,  2014, \mn@doi [\araa]
  {10.1146/annurev-astro-081811-125615}, \href
  {https://ui.adsabs.harvard.edu/abs/2014ARA&A..52..415M} {52, 415}

\bibitem[\protect\citeauthoryear{{Maksym}, {Ulmer}, {Roth}, {Irwin}, {Dupke},
  {Ho}, {Keel}  \& {Adami}}{{Maksym} et~al.}{2014}]{Maksym+14}
{Maksym} W.~P.,  {Ulmer} M.~P.,  {Roth} K.~C.,  {Irwin} J.~A.,  {Dupke} R.,
  {Ho} L.~C.,  {Keel} W.~C.,   {Adami} C.,  2014, \mn@doi [\mnras]
  {10.1093/mnras/stu1485}, \href
  {https://ui.adsabs.harvard.edu/abs/2014MNRAS.444..866M} {444, 866}

\bibitem[\protect\citeauthoryear{{Mallmann} et~al.,}{{Mallmann}
  et~al.}{2018}]{Mallmann+18}
{Mallmann} N.~D.,  et~al., 2018, \mn@doi [\mnras] {10.1093/mnras/sty1364},
  \href {http://adsabs.harvard.edu/abs/2018MNRAS.478.5491M} {478, 5491}

\bibitem[\protect\citeauthoryear{{Martin} et~al.,}{{Martin}
  et~al.}{2007}]{Martin+07}
{Martin} D.~C.,  et~al., 2007, \mn@doi [\apjs] {10.1086/516639}, \href
  {https://ui.adsabs.harvard.edu/abs/2007ApJS..173..342M} {173, 342}

\bibitem[\protect\citeauthoryear{{Martins}, {Riffel}, {Rodr{\'\i}guez-Ardila},
  {Gruenwald}  \& {de Souza}}{{Martins} et~al.}{2010}]{Martins+10}
{Martins} L.~P.,  {Riffel} R.,  {Rodr{\'\i}guez-Ardila} A.,  {Gruenwald} R.,
  {de Souza} R.,  2010, \mn@doi [\mnras] {10.1111/j.1365-2966.2010.16817.x},
  \href {https://ui.adsabs.harvard.edu/abs/2010MNRAS.406.2185M} {406, 2185}

\bibitem[\protect\citeauthoryear{{Menezes}, {Ricci}, {Steiner}, {da Silva},
  {Ferrari}  \& {Borges}}{{Menezes} et~al.}{2019}]{Menezes+19}
{Menezes} R.~B.,  {Ricci} T.~V.,  {Steiner} J.~E.,  {da Silva} P.,  {Ferrari}
  F.,   {Borges} B.~W.,  2019, \mn@doi [\mnras] {10.1093/mnras/sty3337}, \href
  {https://ui.adsabs.harvard.edu/\#abs/2019MNRAS.483.3700M} {483, 3700}

\bibitem[\protect\citeauthoryear{{Mentz} et~al.,}{{Mentz}
  et~al.}{2016}]{Mentz+16}
{Mentz} J.~J.,  et~al., 2016, \mn@doi [\mnras] {10.1093/mnras/stw2129}, \href
  {https://ui.adsabs.harvard.edu/abs/2016MNRAS.463.2819M} {463, 2819}

\bibitem[\protect\citeauthoryear{{Morelli}, {Calvi}, {Masetti}, {Parisi},
  {Landi}, {Maiorano}, {Minniti}  \& {Galaz}}{{Morelli}
  et~al.}{2013}]{Morelli+13}
{Morelli} L.,  {Calvi} V.,  {Masetti} N.,  {Parisi} P.,  {Landi} R.,
  {Maiorano} E.,  {Minniti} D.,   {Galaz} G.,  2013, \mn@doi [\aap]
  {10.1051/0004-6361/201321733}, \href
  {https://ui.adsabs.harvard.edu/abs/2013A&A...556A.135M} {556, A135}

\bibitem[\protect\citeauthoryear{{Mukherjee}, {Bicknell}, {Wagner},
  {Sutherland}  \& {Silk}}{{Mukherjee} et~al.}{2018}]{Mukherjee+18}
{Mukherjee} D.,  {Bicknell} G.~V.,  {Wagner} A.~Y.,  {Sutherland} R.~S.,
  {Silk} J.,  2018, \mn@doi [\mnras] {10.1093/mnras/sty1776}, \href
  {https://ui.adsabs.harvard.edu/abs/2018MNRAS.479.5544M} {479, 5544}

\bibitem[\protect\citeauthoryear{{Oh} et~al.,}{{Oh} et~al.}{2018}]{Oh+18}
{Oh} K.,  et~al., 2018, \mn@doi [\apjs] {10.3847/1538-4365/aaa7fd}, \href
  {https://ui.adsabs.harvard.edu/abs/2018ApJS..235....4O} {235, 4}

\bibitem[\protect\citeauthoryear{{Ramos Almeida} \& {Ricci}}{{Ramos Almeida} \&
  {Ricci}}{2017}]{RamosAlmeida&Ricci17}
{Ramos Almeida} C.,  {Ricci} C.,  2017, \mn@doi [Nature Astronomy]
  {10.1038/s41550-017-0232-z}, \href
  {https://ui.adsabs.harvard.edu/abs/2017NatAs...1..679R} {1, 679}

\bibitem[\protect\citeauthoryear{{Riffel} \& {Storchi-Bergmann}}{{Riffel} \&
  {Storchi-Bergmann}}{2011}]{RiffelRA+11}
{Riffel} R.~A.,  {Storchi-Bergmann} T.,  2011, \mn@doi [\mnras]
  {10.1111/j.1365-2966.2011.19441.x}, \href
  {http://adsabs.harvard.edu/abs/2011MNRAS.417.2752R} {417, 2752}

\bibitem[\protect\citeauthoryear{{Riffel}, {Pastoriza},
  {Rodr{\'{\i}}guez-Ardila}  \& {Bonatto}}{{Riffel} et~al.}{2009}]{Riffel+09}
{Riffel} R.,  {Pastoriza} M.~G.,  {Rodr{\'{\i}}guez-Ardila} A.,   {Bonatto} C.,
   2009, \mn@doi [\mnras] {10.1111/j.1365-2966.2009.15448.x}, \href
  {http://adsabs.harvard.edu/abs/2009MNRAS.400..273R} {400, 273}

\bibitem[\protect\citeauthoryear{{Riffel}, {Storchi-Bergmann}, {Riffel}  \&
  {Pastoriza}}{{Riffel} et~al.}{2010}]{RiffelRA+10}
{Riffel} R.~A.,  {Storchi-Bergmann} T.,  {Riffel} R.,   {Pastoriza} M.~G.,
  2010, \mn@doi [\apj] {10.1088/0004-637X/713/1/469}, \href
  {http://adsabs.harvard.edu/abs/2010ApJ...713..469R} {713, 469}

\bibitem[\protect\citeauthoryear{{Riffel}, {Riffel}, {Ferrari}  \&
  {Storchi-Bergmann}}{{Riffel} et~al.}{2011}]{Riffel+11}
{Riffel} R.,  {Riffel} R.~A.,  {Ferrari} F.,   {Storchi-Bergmann} T.,  2011,
  \mn@doi [\mnras] {10.1111/j.1365-2966.2011.19061.x}, \href
  {http://adsabs.harvard.edu/abs/2011MNRAS.416..493R} {416, 493}

\bibitem[\protect\citeauthoryear{{Riffel}, {Storchi-Bergmann}, {Riffel},
  {Dahmer-Hahn}, {Diniz}, {Sch{\"o}nell}  \& {Dametto}}{{Riffel}
  et~al.}{2017}]{RiffelRA+17}
{Riffel} R.~A.,  {Storchi-Bergmann} T.,  {Riffel} R.,  {Dahmer-Hahn} L.~G.,
  {Diniz} M.~R.,  {Sch{\"o}nell} A.~J.,   {Dametto} N.~Z.,  2017, \mn@doi
  [\mnras] {10.1093/mnras/stx1308}, \href
  {http://adsabs.harvard.edu/abs/2017MNRAS.470..992R} {470, 992}

\bibitem[\protect\citeauthoryear{{Riffel} et~al.,}{{Riffel}
  et~al.}{2018}]{RiffelRA+18}
{Riffel} R.~A.,  et~al., 2018, \mn@doi [\mnras] {10.1093/mnras/stx2857}, \href
  {https://ui.adsabs.harvard.edu/abs/2018MNRAS.474.1373R} {474, 1373}

\bibitem[\protect\citeauthoryear{{Riffel} et~al.,}{{Riffel}
  et~al.}{2021}]{riffel+21}
{Riffel} R.,  et~al., 2021, \mn@doi [\mnras] {10.1093/mnras/staa3907}, \href
  {https://ui.adsabs.harvard.edu/abs/2021MNRAS.501.4064R} {501, 4064}

\bibitem[\protect\citeauthoryear{{Salim} et~al.,}{{Salim}
  et~al.}{2007}]{Salim+07}
{Salim} S.,  et~al., 2007, \mn@doi [\apjs] {10.1086/519218}, \href
  {https://ui.adsabs.harvard.edu/abs/2007ApJS..173..267S} {173, 267}

\bibitem[\protect\citeauthoryear{{S{\'a}nchez} et~al.,}{{S{\'a}nchez}
  et~al.}{2018}]{Sanchez+18}
{S{\'a}nchez} S.~F.,  et~al., 2018, \rmxaa, \href
  {https://ui.adsabs.harvard.edu/abs/2018RMxAA..54..217S} {54, 217}

\bibitem[\protect\citeauthoryear{{Schawinski}, {Koss}, {Berney}  \&
  {Sartori}}{{Schawinski} et~al.}{2015}]{Schawinski+15}
{Schawinski} K.,  {Koss} M.,  {Berney} S.,   {Sartori} L.~F.,  2015, \mn@doi
  [\mnras] {10.1093/mnras/stv1136}, \href
  {https://ui.adsabs.harvard.edu/abs/2015MNRAS.451.2517S} {451, 2517}

\bibitem[\protect\citeauthoryear{{Sch{\"o}nell}, {Storchi-Bergmann}, {Riffel}
  \& {Riffel}}{{Sch{\"o}nell} et~al.}{2017}]{Schonell+17}
{Sch{\"o}nell} Astor~J. J.,  {Storchi-Bergmann} T.,  {Riffel} R.~A.,   {Riffel}
  R.,  2017, \mn@doi [\mnras] {10.1093/mnras/stw2263}, \href
  {https://ui.adsabs.harvard.edu/abs/2017MNRAS.464.1771S} {464, 1771}

\bibitem[\protect\citeauthoryear{{Shimizu} et~al.,}{{Shimizu}
  et~al.}{2019}]{Shimizu+19}
{Shimizu} T.~T.,  et~al., 2019, \mn@doi [\mnras] {10.1093/mnras/stz2802}, \href
  {https://ui.adsabs.harvard.edu/abs/2019MNRAS.490.5860S} {490, 5860}

\bibitem[\protect\citeauthoryear{{Silverman} et~al.,}{{Silverman}
  et~al.}{2008}]{Silverman+08}
{Silverman} J.~D.,  et~al., 2008, \mn@doi [\apj] {10.1086/529572}, \href
  {http://adsabs.harvard.edu/abs/2008ApJ...679..118S} {679, 118}

\bibitem[\protect\citeauthoryear{{Skrutskie} et~al.,}{{Skrutskie}
  et~al.}{2006}]{Skrutskie+06}
{Skrutskie} M.~F.,  et~al., 2006, \mn@doi [\aj] {10.1086/498708}, \href
  {https://ui.adsabs.harvard.edu/abs/2006AJ....131.1163S} {131, 1163}

\bibitem[\protect\citeauthoryear{{Stanley} et~al.,}{{Stanley}
  et~al.}{2017}]{Stanley+17}
{Stanley} F.,  et~al., 2017, \mn@doi [\mnras] {10.1093/mnras/stx2121}, \href
  {http://adsabs.harvard.edu/abs/2017MNRAS.472.2221S} {472, 2221}

\bibitem[\protect\citeauthoryear{{Storchi-Bergmann}, {Raimann}, {Bica}  \&
  {Fraquelli}}{{Storchi-Bergmann} et~al.}{2000}]{StorchiBergmann+00}
{Storchi-Bergmann} T.,  {Raimann} D.,  {Bica} E.~L.~D.,   {Fraquelli} H.~A.,
  2000, \mn@doi [\apj] {10.1086/317247}, \href
  {http://adsabs.harvard.edu/abs/2000ApJ...544..747S} {544, 747}

\bibitem[\protect\citeauthoryear{{Storchi-Bergmann}, {Gonz{\'a}lez Delgado},
  {Schmitt}, {Cid Fernandes}  \& {Heckman}}{{Storchi-Bergmann}
  et~al.}{2001}]{StorchiBergmann+01}
{Storchi-Bergmann} T.,  {Gonz{\'a}lez Delgado} R.~M.,  {Schmitt} H.~R.,  {Cid
  Fernandes} R.,   {Heckman} T.,  2001, \mn@doi [\apj] {10.1086/322290}, \href
  {https://ui.adsabs.harvard.edu/abs/2001ApJ...559..147S} {559, 147}

\bibitem[\protect\citeauthoryear{{Storchi-Bergmann}, {Riffel}, {Riffel},
  {Diniz}, {Borges Vale}  \& {McGregor}}{{Storchi-Bergmann}
  et~al.}{2012}]{StorchiBergmann+12}
{Storchi-Bergmann} T.,  {Riffel} R.~A.,  {Riffel} R.,  {Diniz} M.~R.,  {Borges
  Vale} T.,   {McGregor} P.~J.,  2012, \mn@doi [\apj]
  {10.1088/0004-637X/755/2/87}, \href
  {http://adsabs.harvard.edu/abs/2012ApJ...755...87S} {755, 87}

\bibitem[\protect\citeauthoryear{{Vazdekis}, {S{\'a}nchez-Bl{\'a}zquez},
  {Falc{\'o}n-Barroso}, {Cenarro}, {Beasley}, {Cardiel}, {Gorgas}  \&
  {Peletier}}{{Vazdekis} et~al.}{2010}]{Vazdekis+10}
{Vazdekis} A.,  {S{\'a}nchez-Bl{\'a}zquez} P.,  {Falc{\'o}n-Barroso} J.,
  {Cenarro} A.~J.,  {Beasley} M.~A.,  {Cardiel} N.,  {Gorgas} J.,   {Peletier}
  R.~F.,  2010, \mn@doi [\mnras] {10.1111/j.1365-2966.2010.16407.x}, \href
  {https://ui.adsabs.harvard.edu/abs/2010MNRAS.404.1639V} {404, 1639}

\bibitem[\protect\citeauthoryear{{Wilkinson} et~al.,}{{Wilkinson}
  et~al.}{2015}]{Wilkinson+15}
{Wilkinson} D.~M.,  et~al., 2015, \mn@doi [\mnras] {10.1093/mnras/stv301},
  \href {https://ui.adsabs.harvard.edu/abs/2015MNRAS.449..328W} {449, 328}

\bibitem[\protect\citeauthoryear{{Worthey}}{{Worthey}}{1994}]{Worthey94}
{Worthey} G.,  1994, \mn@doi [\apjs] {10.1086/192096}, \href
  {http://adsabs.harvard.edu/abs/1994ApJS...95..107W} {95, 107}

\bibitem[\protect\citeauthoryear{{Wyder} et~al.,}{{Wyder}
  et~al.}{2007}]{Wyder+07}
{Wyder} T.~K.,  et~al., 2007, \mn@doi [\apjs] {10.1086/521402}, \href
  {https://ui.adsabs.harvard.edu/abs/2007ApJS..173..293W} {173, 293}

\bibitem[\protect\citeauthoryear{{Xilouris}, {Madden}, {Galliano}, {Vigroux}
  \& {Sauvage}}{{Xilouris} et~al.}{2004}]{Xilouris+04}
{Xilouris} E.~M.,  {Madden} S.~C.,  {Galliano} F.,  {Vigroux} L.,   {Sauvage}
  M.,  2004, \mn@doi [\aap] {10.1051/0004-6361:20034020}, \href
  {http://adsabs.harvard.edu/abs/2004A%26A...416...41X} {416, 41}

\bibitem[\protect\citeauthoryear{{Zier} \& {Biermann}}{{Zier} \&
  {Biermann}}{2002}]{Zier&Biermann02}
{Zier} C.,  {Biermann} P.~L.,  2002, \mn@doi [\aap]
  {10.1051/0004-6361:20021339}, \href
  {http://adsabs.harvard.edu/abs/2002A%26A...396...91Z} {396, 91}

\bibitem[\protect\citeauthoryear{{Zubovas} \& {Bourne}}{{Zubovas} \&
  {Bourne}}{2017}]{Zubovas&Bourne17}
{Zubovas} K.,  {Bourne} M.~A.,  2017, \mn@doi [\mnras] {10.1093/mnras/stx787},
  \href {http://adsabs.harvard.edu/abs/2017MNRAS.468.4956Z} {468, 4956}

\bibitem[\protect\citeauthoryear{{de Vaucouleurs}, {de Vaucouleurs}, {Corwin},
  {Buta}, {Paturel}  \& {Fouqu{\'e}}}{{de Vaucouleurs}
  et~al.}{1991}]{deVaucouleurs+91}
{de Vaucouleurs} G.,  {de Vaucouleurs} A.,  {Corwin} Jr. H.~G.,  {Buta} R.~J.,
  {Paturel} G.,   {Fouqu{\'e}} P.,  1991, {Third Reference Catalogue of Bright
  Galaxies. Volume I: Explanations and references. Volume II: Data for galaxies
  between 0$^{h}$ and 12$^{h}$. Volume III: Data for galaxies between 12$^{h}$
  and 24$^{h}$.}

\bibitem[\protect\citeauthoryear{{van den Bosch}}{{van den
  Bosch}}{2016}]{vandenBosch16}
{van den Bosch} R. C.~E.,  2016, \mn@doi [\apj] {10.3847/0004-637X/831/2/134},
  \href {https://ui.adsabs.harvard.edu/abs/2016ApJ...831..134V} {831, 134}

\makeatother
\end{thebibliography}

\bsp
\label{lastpage}

\end{document}